\documentclass[jpb,twocolumn,amsmath,amssymb]{revtex4}

\usepackage{graphicx,epsfig}
\usepackage{color}

\begin{document}

\newcommand{\SeeLL}{\mathcal{S}^{ee}_{LL}} 
\newcommand{\SehLL}{\mathcal{S}^{eh}_{LL}}
\newcommand{\SehRR}{\mathcal{S}^{eh}_{RR}} 
\newcommand{\SheLL}{\mathcal{S}^{he}_{LL}} 
\newcommand{\ShhLL}{\mathcal{S}^{hh}_{LL}} 
\newcommand{\ShhRR}{\mathcal{S}^{hh}_{RR}} 
\newcommand{\SeeLR}{\mathcal{S}^{ee}_{LR}} 
\newcommand{\SehLR}{\mathcal{S}^{eh}_{LR}} 
\newcommand{\SheLR}{\mathcal{S}^{he}_{LR}} 
\newcommand{\ShhLR}{\mathcal{S}^{hh}_{LR}} 
\newcommand{\Seepq}{\mathcal{S}^{ee}_{pq}} 
\newcommand{\Sehpq}{\mathcal{S}^{eh}_{pq}} 
\newcommand{\Shepq}{\mathcal{S}^{he}_{pq}}
\newcommand{\Shhpq}{\mathcal{S}^{hh}_{pq}} 
\newcommand{\rLL}{r_{LL}}
\newcommand{\rRR}{r_{RR}}
\newcommand{\rLR}{r_{LR}}
\newcommand{\rRL}{r_{RL}}
\newcommand{\rpq}{r_{pq}}

\title{Mixed, charge and heat noises in thermoelectric nanosystems}

\author{Adeline Cr\'epieux$^1$}
\author{Fabienne Michelini$^2$}
\affiliation{$^1$Aix Marseille Universit\'e, Universit\'e de Toulon, CNRS, CPT UMR 7332, 13288 Marseille, France}
\affiliation{$^2$Aix Marseille Universit\'e, CNRS, IM2NP, UMR 7334, 13288 Marseille, France}

\begin{abstract}

Mixed, charge and heat current fluctuations as well as thermoelectric differential conductances are considered for non-interacting nanosystems connected to reservoirs. Using the Landauer-B\"uttiker formalism, we derive general expressions for these quantities and consider their possible relationships in the entire ranges of temperature, voltage and  coupling to environment or reservoirs. We introduce a dimensionless quantity given by the ratio between the product of mixed noises and the product of charge and heat noises, distinguishing between the auto-ratio defined in the same reservoir and the cross-ratio between distinct reservoirs. From the linear response regime to the high voltage regime, we further specify the analytical expressions of differential conductances, noises and ratios of noises, and examine their behavior in two concrete nanosystems: a quantum point contact in an ohmic environment and a single energy level quantum dot connected to reservoirs. In the linear response regime, we find that these ratios are equal to each other and are simply related to the figure of merit. They can also be expressed in terms of differential conductances with the help of the fluctuation-dissipation theorem. In the non-linear regime, these ratios radically distinguish between themselves as the auto-ratio remains bounded by one while the cross-ratio exhibits rich and complex behaviors. In the quantum dot nanosystem, we moreover demonstrate that the thermoelectric efficiency can be expressed as a ratio of noises in the non-linear Schottky regime. In the intermediate voltage regime, the cross-ratio changes sign and diverges, which evidences a change of sign in the heat cross-noise.
\end{abstract}

\maketitle
%%%%%%%%%%%%%%%%%%%%%%%%%%%%%%%%%%%%%%%%%%%%%%%%%%%%%%%%%%%%%%%%%%
%																 %
%																 %
%		INTRODUCTION											 %
%																 %
%																 %
%%%%%%%%%%%%%%%%%%%%%%%%%%%%%%%%%%%%%%%%%%%%%%%%%%%%%%%%%%%%%%%%%%

\section{Introduction} 

The main motivation for studying thermoelectricity in quantum systems is the promise to increase the conversion efficiency by reducing the dimension. Indeed, the first measurements of values larger than one for the figure of merit were obtained in superlattices and quantum dot superlattices \cite{venkata01,harman02}. These observations were the trigger signal of a large number of both experimental and theoretical works. An increase of the thermopower has been obtained in molecular junctions between a gold substrate and a gold scanning tunneling microscope tip~\cite{reddy07}, and measurements of the Seebeck coefficient have been performed in carbon nanotubes \cite{sumanasekera02,small03}, control break junctions~\cite{ludoph99}, magnetic tunnel junctions \cite{walter11}, spin valves~\cite{bakker10}, and Kondo quantum dots~\cite{scheibner05}. Non-linear thermovoltage and thermocurrent in quantum dot have been highlighted~\cite{fahlvik13}.

Extensive theoretical works were performed in order to understand how thermoelectric properties are affected in nanosystems with  multi channels \cite{sivan86}, multi-terminals \cite{butcher90}, on-site interaction \cite{azema12,dutt13}, inelastic scattering \cite{matveev02,entin10}, and time-dependent voltage \cite{arrachea07,crepieux11}. In addition, the validity of the Onsager relation linking the Seebeck and Peltier coefficients, the validity of the Wiedemann-Franz and Fourier laws were questioned in nanosystems \cite{butcher90,iyoda10,vavilov05,dubi09}. Indeed, the breakdown of thermoelectric reciprocity relations has been experimentally observed recently in a four-terminals mesoscopic device \cite{matthews13}. A key point for thermoelectricity in nanosystems is the fact that the tools used to quantify the efficiency for classical systems fail to describe the quantum ones. In particular, the figure of merit is a concept which makes sense only in the linear response regime. Indeed, the optimization of the figure of merit outside the linear response regime does not guarantee the optimization of the efficiency \cite{muralidharan12}. The appropriate approach is to rather consider directly the efficiency, and look for the optimization of the ratio between electrical and heat powers \cite{whitney13,whitney14,kennes13}.

In parallel, the interest in heat noise in quantum systems is growing up \cite{krive01,kindermann04,zhan11,sergi11,sanchez12,battista13}. However, these studies are mainly restricted to correlations between the heat current and itself. In particular, it has been shown that the correlator between heat currents in distinct reservoirs is not necessary negative contrary to what happens with charge currents \cite{moskalets14}.  It is only recently that the correlation between heat and charge currents -- what we call mixed noise -- has been considered for thermoelectricity \cite{giazotto06}. In particular, to analyse a quantum-dot based engine, Sanchez and co-workers introduced a ratio between the different kinds of noises which is maximal for the optimal configuration of this particular thermoelectric nanosystem \cite{sanchez13}.

In this work, we adopt a general viewpoint and investigate in great details the mixed noise and ratios of noises using the  Landauer-B\"uttiker formalism \cite{blanter00}. We derived the explicit expressions of heat, charge and mixed noises as well as ratios of noises and thermoelectric differential conductances for the following two nanosystems: a quantum point contact (QPC) coupled to an ohmic environment and a quantum dot (QD) connected to two reservoirs.

In the linear response regime, we recover fluctuation-dissipation theorems which link the electrical and thermal conductances to the charge and heat noises. The same kind of fluctuation-dissipation theorem applies for mixed noises provided that one considers the mixed thermoelectric conductances. As a consequence, the figure of merit is related to the ratio between the square of mixed noise and the product of heat and charge noises. 
In the non-linear response regime, we were guided to distinguish between two ratios of noises: the cross-ratio that is defined between two different reservoirs and the auto-ratio defined inside the same reservoir. Significantly, for the two nanosystems, the cross-ratio can reach larger values than one, while the auto-ratio cannot exceed one. In between these two regimes, the cross-ratio reveals complex behaviors that we connect to the features of the different kind of noises.  Despite this complexity, we find that in the Schottky regime the thermoelectric efficiency is still given by the noises but with an expression which differs from the one obtained in the linear response regime.

The paper is organized as follows: in Sec.~II, we define all the quantities we are interested in, i.e., differential conductances, current noises at zero frequency and ratios of noises. We give the general expressions of these quantities obtained in the framework of the Landauer-B\"uttiker scattering theory in Sec.~III, and their reduced expressions in both the linear response regime and high voltage regime in Sec.~IV. In Secs.~V and VI, we apply our results to a QPC and a QD and discuss various regimes. We conclude in Sec.~VII.

%%%%%%%%%%%%%%%%%%%%%%%%%%%%%%%%%%%%%%%%%%%%%%%%%%%%%%%%%%%%%%%%%%
%																 %
%																 %
%		DEFINITIONS											 %
%																 %
%																 %
%%%%%%%%%%%%%%%%%%%%%%%%%%%%%%%%%%%%%%%%%%%%%%%%%%%%%%%%%%%%%%%%%%

\section{Definitions}\label{definitions}

We define the following zero-frequency current correlators between the reservoirs $p$ and $q$:
\begin{eqnarray}
\mathcal{S}^{\alpha\beta}_{pq}&=&\int \langle\delta \hat I^\alpha_{p}(t)\delta\hat I^{\beta}_{q}(0)\rangle dt~,
\end{eqnarray}
where $\delta\hat I^{e,h}_{p}(t)=\hat I^{e,h}_{p}(t)-I^{e,h}_{p}$, with $\hat{I}^{e(h)}_p$ the charge (heat) current operator, and $I^{e(h)}_p=\langle \hat{I}^{e(h)}_p\rangle$ the charge (heat) average current in the reservoir $p$. $\Seepq$ corresponds to the charge noise and $\Shhpq$ corresponds to the heat noise, whereas $\Sehpq$ and $\Shepq$ correspond to the correlations between charge and heat currents. We call them \emph{mixed} noises. In the following, we restrict our work to a two terminal system, thus $\{p,q\}\in\{L,R\}$, where $L$ ($R$) refers to the left (right) reservoir driven at chemical potential $\mu_{L(R)}$ and temperature $T_{L(R)}$. 
As a general rule, we call \emph{auto}-quantities when calculated for $p=q$, and \emph{cross}-quantities when calculated for $p\neq q$.

Next, we introduce the differential conductances defined as:
\begin{eqnarray}
G_{p}&=&e\frac{\partial I^e_{p}}{\partial \mu_{p}}~,\;\;\;
K_{p}=\frac{\partial I^h_{p}}{\partial T_{p}}~,\\
X_{p}&=&\frac{\partial I^e_{p}}{\partial T_{p}}~,\;\;\;
Y_{p}=e\frac{\partial I^h_{p}}{\partial \mu_{p}}~.
\end{eqnarray}
$G_{p}$ and $K_{p}$ correspond to the electrical and thermal differential conductances, whereas $X_{p}$ and $Y_{p}$ are  differential mixed conductances that locally reflect the thermoelectric conversion. In the linear response regime, these last two conductances are related to the Seebeck and Peltier coefficients (see Sec.~\ref{high_temp}). In the non-linear response regime, these differential conductances are the adequate quantities to consider since the currents vary as power laws with the voltage as we will show in Secs.~V and VI. We also use average conductances merely defined as: $G=(G_L+G_R)/2$, $X=(X_L+X_R)/2$, $Y=(Y_L+Y_R)/2$, and $K=(K_L+K_R)/2$.

Finally, we introduce a dimensionless quantity: the ratio between the product of mixed correlations on the one hand, and the product of charge and heat ones on the other hand:
\begin{eqnarray}
\label{def_ratio_noise}
\rpq&=&\frac{\Sehpq\Shepq}{ \Seepq\Shhpq}~.
\end{eqnarray}
This ratio gives indications on the mixed correlations between the heat and charge currents: (i) $\rpq=0$ when heat and charge currents are uncorrelated between reservoirs $p$ and $q$, (ii) $r_{pp}=1$ when heat and charge currents are  maximally correlated and, (iii) $r_{p\ne q}\rightarrow\infty$ when the left and right heat currents are uncorrelated.  Indeed, whereas $|r_{pp}|\leq 1$ because of the Cauchy-Schwarz inequality, we will obtain here that no such limitation applies for $r_{p\ne q}$. In addition, the fact that the cross-ratio and auto-ratio differ or that $r_{p\ne q}>1$ means that the system operates outside the linear response regime. Information about the mixed correlation between energy and charge currents can be obtained from this ratio. Indeed we have 
equivalently:
\begin{eqnarray}\label{energy}
r_{pq}=
1+\frac{\mathcal{S}^{eE}_{pq}\mathcal{S}^{Ee}_{pq}-\mathcal{S}^{ee}_{pq}\mathcal{S}^{EE}_{pq}}{\Seepq\Shhpq}~,
\end{eqnarray}
where $\mathcal{S}^{EE}_{pq}$ is the energy noise and $\mathcal{S}^{eE(Ee)}_{pq}$ is the charge-energy noise which measure the correlations related to the energy current: $I^E_p=I^h_p+(\mu_p/e)I^e_p$. Having  $r_{pq}=1$ means either that energy fluctuations are absent: $\mathcal{S}^{eE}_{pq}=\mathcal{S}^{Ee}_{pq}=S_{pq}^{EE}=0$ ($\mathcal{S}^{ee}_{pq}$ never cancels at finite temperature and/or finite voltage), or that there is an exact compensation between the charge-energy and energy-energy correlators, i.e.,  $\mathcal{S}^{eE}_{pq}\mathcal{S}^{Ee}_{pq}=\mathcal{S}^{ee}_{pq}\mathcal{S}^{EE}_{pq}$. Note that another type of ratio was defined in Ref.~\onlinecite{sanchez13} for a three-terminal quantum dot engine that could be written as $r=(\mathcal{S}^{eh}_{13})^2/(\mathcal{S}^{ee}_{11}\mathcal{S}^{hh}_{33})$. It is still bounded and it reaches the value one when mixed correlations are maximal.

%%%%%%%%%%%%%%%%%%%%%%%%%%%%%%%%%%%%%%%%%%%%%%%%%%%%%%%%%%%%%%%%%%
%																 %
%																 %
%		RESULTS AND GENERAL RELATIONS										 %
%																 %
%																 %
%%%%%%%%%%%%%%%%%%%%%%%%%%%%%%%%%%%%%%%%%%%%%%%%%%%%%%%%%%%%%%%%%%

\section{Landauer-\textit{like} expressions}

We derive the formal expressions of the differential conductances, and heat, charge and mixed noises at zero frequency within the Landauer-B\"uttiker scattering theory \cite{blanter00}. We assume that the transmission coefficient $\mathcal{T}$ through the nanoscopic conductor does not depend on the external variables $\mu_{L,R}$ and $T_{L,R}$ \cite{note1}.

%%%%%%%%%%%%%%%%%%%
\subsection{Differential conductances}

To get the differential conductances, we use the Landauer expressions of charge and heat average currents:
\begin{eqnarray}\label{def_curr_elec}
I^e_{L,R}&=&\pm\frac{e}{h}\int^{\infty}_{-\infty}\left[f_L(\epsilon)-f_R(\epsilon)\right]\mathcal{T}(\epsilon)d\epsilon~,
\\\label{def_curr_ther}
I^h_{L,R}&=&\pm\frac{1}{h}\int^{\infty}_{-\infty}(\epsilon-\mu_{L,R})\left[f_L(\epsilon)-f_R(\epsilon)\right]\mathcal{T}(\epsilon)d\epsilon~,\nonumber\\
\end{eqnarray}
where $f_p(\epsilon)=\left[1+e^{(\epsilon-\mu_p)/(k_B T_p)}\right]^{-1}$ is the Fermi-Dirac distribution function and the sign $+(-)$ holds for reservoir L(R). The convention chosen for the current directions is to consider the flux of electrons or heat from the reservoirs to the central part of the system.
The calculation of their derivatives according to $\mu_{p}$ or $T_{p}$ leads to:
\begin{eqnarray}\label{exp_cond_elec}
G_{p}&=&\frac{e^2}{hk_BT_{p}}\int^{\infty}_{-\infty}f_{p}(\epsilon)[1-f_{p}(\epsilon)]\mathcal{T}(\epsilon)d\epsilon~,
\end{eqnarray}
\begin{eqnarray}\label{exp_cond_x}
X_{p}&=&\frac{e}{hk_BT^2_{p}} \int^{\infty}_{-\infty}(\epsilon-\mu_{p})f_{p}(\epsilon)[1-f_{p}(\epsilon)]\mathcal{T}(\epsilon)d\epsilon~,\nonumber\\
\end{eqnarray}
\begin{eqnarray}\label{exp_cond_y}
Y_{p}&=&-I^e_{p}+\frac{e}{hk_BT_{p}}\int^{\infty}_{-\infty}(\epsilon-\mu_{p})\nonumber\\
&&\times f_{p}(\epsilon)[1-f_{p}(\epsilon)]\mathcal{T}(\epsilon)d\epsilon~,
\end{eqnarray}
and,
\begin{eqnarray}\label{exp_cond_ther}
K_{p}&=&\frac{1}{hk_BT^2_{p}}\int^{\infty}_{-\infty}(\epsilon-\mu_{p})^2
f_{p}(\epsilon)[1-f_{p}(\epsilon)]\mathcal{T}(\epsilon)d\epsilon~.\nonumber\\
\end{eqnarray}
These conductances obey the relation: $Y_{p}=T_{p}X_{p}-I^e_{p}$, which reduces to $Y_{p}=T_{p}X_{p}$ in the linear response regime (Onsager relation).

%%%%%%%%%%%%%%
\subsection{Current noises}

Within the Landauer-B\"uttiker scattering theory, the zero-frequency charge, mixed and heat noises are given by:
\begin{eqnarray}
\label{def_charge_noise}
\Seepq&=&(2\delta_{pq}-1)\frac{e^2}{h}\int^{\infty}_{-\infty}\mathcal{F}(\epsilon)d\epsilon~,
\end{eqnarray}
\begin{eqnarray}\label{def_mixte_noise_1}
\Sehpq&=&(2\delta_{pq}-1)\frac{e}{h}\int^{\infty}_{-\infty}(\epsilon-\mu_{q})\mathcal{F}(\epsilon)d\epsilon~,
\end{eqnarray}
\begin{eqnarray}\label{def_mixte_noise_2}
\Shepq&=&(2\delta_{pq}-1)\frac{e}{h}\int^{\infty}_{-\infty}(\epsilon-\mu_{p})\mathcal{F}(\epsilon)d\epsilon~,
\end{eqnarray}
and,
\begin{eqnarray}\label{def_heat_noise}
\Shhpq&=&(2\delta_{pq}-1)\frac{1}{h}\int^{\infty}_{-\infty}(\epsilon-\mu_{p})(\epsilon-\mu_{q})\mathcal{F}(\epsilon)d\epsilon~,\nonumber\\
\end{eqnarray}
where the factor $(2\delta_{pq}-1)$ gives $1$ when $p=q$, or $-1$ when $p\ne q$, and
\begin{eqnarray}
\mathcal{F}(\epsilon)&=&\mathcal{T}(\epsilon)\Big[f_L(\epsilon)[1-f_L(\epsilon)]+f_R(\epsilon)[1-f_R(\epsilon)]\Big]\nonumber\\
&&+\mathcal{T}(\epsilon)[1-\mathcal{T}(\epsilon)][f_L(\epsilon)-f_R(\epsilon)]^2~.
\end{eqnarray}
These correlators are connected to each other. For the charge noises, we have:
$\mathcal{S}^{ee}_{pq}=\mathcal{S}^{ee}_{qp}$, $\mathcal{S}^{ee}_{pp}=\mathcal{S}^{ee}_{\bar p \bar p}$, and $\mathcal{S}^{ee}_{p\bar p}=-\mathcal{S}^{ee}_{pp}$, where $\bar p=L$ when $p=R$, and $\bar p=R$ when $p=L$.
For the mixed noises, we have $\mathcal{S}^{eh}_{pq}=\mathcal{S}^{he}_{qp}$, and,
\begin{eqnarray}
&& \Sehpq=\Shepq+(\mu_p-\mu_q)\Seepq/e~,\\
&&\mathcal{S}^{eh}_{pp}=\mathcal{S}^{eh}_{\bar p \bar p}+(\mu_{\bar p}-\mu_p)\mathcal{S}^{ee}_{pp}/e~,\\
&&\mathcal{S}^{eh}_{p \bar p}=- \mathcal{S}^{eh}_{pp}+(\mu_{\bar p}-\mu_p)\mathcal{S}^{ee}_{pp}/e~,
\end{eqnarray}
which reduce to  $\Sehpq=\Shepq=\mathcal{S}^{he}_{qp}$, $\mathcal{S}^{eh}_{pp}=\mathcal{S}^{eh}_{\bar p \bar p}$ and $\mathcal{S}^{eh}_{p \bar p}=-\mathcal{S}^{eh}_{pp}$ in the linear response regime.
For the heat noises, we have $\mathcal{S}^{hh}_{pq}=\mathcal{S}^{hh}_{qp}$ and,
\begin{eqnarray}
&&\mathcal{S}^{hh}_{pp}=\mathcal{S}^{hh}_{\bar p\bar p}+2(\mu_{\bar p} - \mu_p)\mathcal{S}^{eh}_{\bar p\bar p}/e+(\mu_{\bar p}-\mu_p)^2\mathcal{S}^{ee}_{\bar p\bar p}/e^2~,\nonumber\\
\end{eqnarray}
which reduces to $\mathcal{S}^{hh}_{pp}=\mathcal{S}^{hh}_{\bar p\bar p}$ in the linear response regime. As a major consequence of these relations, we conclude that the cross noises and the cross-ratio defined by Eq.~(\ref{def_ratio_noise}) in between the two reservoirs is symmetric when we interchange the reservoirs: $\mathcal{S}^{\alpha\beta}_{RL}=\mathcal{S}^{\beta\alpha}_{LR}$ and $\rRL=\rLR$, thus we will discuss neither $\mathcal{S}^{\alpha\beta}_{RL}$ nor $\rRL$ in the paper. Inversely, $\rLL$ and $\rRR$ may differ as will be the case for the QD nanosystem (see Sec.~VI). 
Moreover, combining these relations, we deduce:
\begin{eqnarray}\label{Seetot}
\sum_{p,q\in\{L,R\}}S^{ee}_{pq}&=&0~,\\\label{Sehtot}
\sum_{p,q\in\{L,R\}}S^{eh}_{pq}&=&0~,\\
\label{Shhtot}
\sum_{p,q\in\{L,R\}}S^{hh}_{pq}&=&(\mu_{L}-\mu_R)^2S^{ee}_{LL}/e^2~.
\end{eqnarray}
In the limit of zero voltage, the total heat noise cancels in agreement with Ref.~\onlinecite{sergi11}.
 The total charge noise, given by Eq.~(\ref{Seetot}), is equal to zero due to charge current fluctuations conservation, whereas the total heat noise, given by Eq.~(\ref{Shhtot}), is equal to the product of the bias voltage square by the charge auto-correlator. It corresponds to a conservation of power fluctuations since it leads to an equality between the thermal power fluctuations and the electric power fluctuations:
\begin{eqnarray}
\int^{\infty}_{-\infty} \langle P^\mathrm{th}(t)P^\mathrm{th}(0)\rangle dt=\int^{\infty}_{-\infty}  \langle P^\mathrm{el}(t)P^\mathrm{el}(0)\rangle dt~,
\end{eqnarray}
where $P^\mathrm{th}=I_L^h+I_R^h$ is the thermal power and $P^\mathrm{el}=VI_L^e$, the electrical power.

%%%%%%%%%%%%%%%%%%%%%%%%%%%%%%%%%%
\subsection{Relations between noises and differential conductances}

We want to express noises in terms of differential conductances defined in Sec.~\ref{definitions}.
Reporting Eqs.~(\ref{exp_cond_elec}) to (\ref{exp_cond_ther}) in the expressions of the charge, mixed and heat correlators given by Eqs.~(\ref{def_charge_noise}) to (\ref{def_heat_noise}), we get:
\begin{widetext}
\begin{eqnarray}
\label{charge_noise}
\Seepq&=&(2\delta_{pq}-1)\Bigg[k_BT_pG_p+k_BT_{\bar p}G_{\bar p}
+\frac{e^2}{h}\int^{\infty}_{-\infty}\mathcal{T}(\epsilon)[1-\mathcal{T}(\epsilon)][f_p(\epsilon)-f_{\bar p}(\epsilon)]^2d\epsilon\Bigg]~,
\end{eqnarray}
\begin{eqnarray}
\label{mixed_noise_1}
\Sehpq&=&(2\delta_{pq}-1)\Bigg[k_BT_{p}^2X_{p}+k_BT_{\bar p}^2X_{\bar p}+(\mu_{\bar q}-\mu_q)k_BT_{\bar q}G_{\bar q}/e+\frac{e}{h}\int^{\infty}_{-\infty}(\epsilon-\mu_{q})\mathcal{T}(\epsilon)[1-\mathcal{T}(\epsilon)][f_p(\epsilon)-f_{\bar p}(\epsilon)]^2d\epsilon\Bigg]~,\nonumber\\
\end{eqnarray}
\begin{eqnarray}
\label{mixed_noise_2}
\Shepq&=&(2\delta_{pq}-1)\Bigg[k_BT_{p}^2X_{p}+k_BT_{\bar p}^2X_{\bar p}+(\mu_{\bar p}-\mu_p)k_BT_{\bar p}G_{\bar p}/e+\frac{e}{h}\int^{\infty}_{-\infty}(\epsilon-\mu_{p})\mathcal{T}(\epsilon)[1-\mathcal{T}(\epsilon)][f_p(\epsilon)-f_{\bar p}(\epsilon)]^2d\epsilon\Bigg]~,\nonumber\\
\end{eqnarray}
and,
\begin{eqnarray}
\label{heat_noise}
\Shhpq&=&(2\delta_{pq}-1)\Bigg[k_BT^2_{p}K_{p}+k_BT^2_{\bar p}K_{\bar p}+(\mu_p-\mu_{\bar q})^2k_BT_{\bar p}G_{\bar p}/e^2+(\mu_{\bar p}-\mu_p)k_BT_{\bar p}^2X_{\bar p}/e+(\mu_{\bar q}-\mu_q)k_BT_{\bar q}^2X_{\bar q}/e\nonumber\\
&&+\frac{1}{h}\int^{\infty}_{-\infty}(\epsilon-\mu_{p})(\epsilon-\mu_{q})\mathcal{T}(\epsilon)[1-\mathcal{T}(\epsilon)][f_p(\epsilon)-f_{\bar p}(\epsilon)]^2d\epsilon\Bigg]~.
\end{eqnarray}
\end{widetext}
With the help of these relations, we discuss in the next section the behavior of the different types of noise in two extreme regimes.

%%%%%%%%%%%%%%%%%%%%%%%%%%%%%%%%%%%%%%%%%%%%%%%%%%%%%%%%%%%%%%%%%%
%																 %
%																 %
%		VARIOUS REGIMES										 %
%																 %
%																 %
%%%%%%%%%%%%%%%%%%%%%%%%%%%%%%%%%%%%%%%%%%%%%%%%%%%%%%%%%%%%%%%%%%

\section{Linear response regime and high voltage regime}

Now, we specify $\mu_{L,R}$ and $T_{L,R}$ in terms of voltage gradient $V$ and temperature gradient $T$ between the reservoirs: $\mu_{L,R}=\epsilon_F\pm eV/2$ and $T_{L,R}=T_0\pm T/2$, where $\epsilon_F$ is the Fermi energy of the reservoirs (set to zero in the following) and $T_0$ is the average temperature of the system.

%%%%%%%%%%%%%%%%%%%%%%%%%%%%%%%%%%%%%
\subsection{Linear response regime}\label{high_temp}

In this regime we have $\{eV,k_BT\}\ll k_BT_0$, thus all terms except the first two are negligible in the right-hand sides of Eqs.~(\ref{charge_noise}) to (\ref{heat_noise}),  and we can write the noises in terms of conductances and average temperature: 
\begin{eqnarray}\label{See_linear}
\mathcal{S}^{ee}_{LL}&=&\mathcal{S}^{ee}_{RR}=2k_BT_0G=-\mathcal{S}^{ee}_{LR}~,\\\label{Seh_linear}
\mathcal{S}^{eh}_{LL}&=&\mathcal{S}^{eh}_{RR}=2k_BT_0^2X=-\mathcal{S}^{he}_{LR}~,\\\label{She_linear}
\mathcal{S}^{he}_{LL}&=&\mathcal{S}^{he}_{RR}=2k_BT_0Y=-\mathcal{S}^{eh}_{LR}~,\\\label{Shh_linear}
\mathcal{S}^{hh}_{LL}&=&\mathcal{S}^{hh}_{RR}=2k_BT_0^2K=-\mathcal{S}^{hh}_{LR}~.
\end{eqnarray}
We note that the auto- and cross-correlations have the same absolute value. Equations~(\ref{See_linear}) and (\ref{Shh_linear}) correspond to the fluctuation-dissipation theorem for charge and heat noises respectively. There exists also direct links between the mixed noises and the thermoelectric conductances given by Eqs.~(\ref{Seh_linear}) and (\ref{She_linear}) in agreement with Ref.~\onlinecite{kubo57}. In addition, we have $Y=XT_0$ (Onsager relation).

From Eqs.~(\ref{See_linear}) to (\ref{Shh_linear}), we directly deduced that all auto- and cross-ratios are identical in the linear response limit and given by the ratio of conductances:
\begin{eqnarray}\label{ratioss_linear}
\rLL=\rRR=\rLR=\frac{XY}{GK}~.
\end{eqnarray}
In addition, it can been shown that $X$ and $Y$ are related to the Seebeck $S$ and Peltier $\Pi$ coefficients through the relations:
\begin{eqnarray}\label{X_linear}
X&=& -GS~,\\\label{Y_linear}
Y&=&-\Pi G=-ST_0G~.
\end{eqnarray}
With the help of these results, the thermoelectric figure of merit defined as $ZT_0=S^2T_0G/\tilde K$, where $\tilde K=\partial I^h/\partial T|_{I^e=0}=K-\Pi S G$ is the thermal conductance at zero charge current, can be fully expressed either in terms of conductances, or in terms of noises. Indeed, from Eqs.~(\ref{See_linear}) to (\ref{Y_linear}), we get:
\begin{eqnarray}\label{ZT0}
ZT_0=\frac{XY}{G\tilde K}=\frac{S^{eh}_{pq}S^{he}_{pq}}{S^{ee}_{pq}S^{hh}_{pq}-S^{eh}_{pq}S^{he}_{pq}}=\frac{\rpq}{1-\rpq}~,
\end{eqnarray}
which is verified whatever the choice of the reservoirs $p$ and $q$. Thus, in the linear response regime, the figure of merit for thermoelectricity is given by the ratio between the product of mixed noises and the product of heat and charge noises. This ratio is hence relevant to quantify the efficiency of the thermoelectric conversion. Because of the Cauchy-Swartz inequality, we have $S^{eh}_{pp}S^{he}_{pp}\le S^{ee}_{pp}S^{hh}_{pp}$. 
As a major consequence, the value taken by $|r_{pp}|$ is contained in the interval $[0,1]$ which implies through Eq.~(\ref{ZT0}) that $ZT_0$ is not bounded. This result is in agreement with Littman and Davidson \cite{littman61} who have instead used an argument of entropy production in their demonstration. 
We note that if $ZT_0=S^2T_0G/\tilde K$ appears as the relevant parameter when the efficiency is maximized according to the charge current since $\eta_\mathrm{max}=\eta_C(\sqrt{1+ZT_0}-1)/(\sqrt{1+ZT_0}+1)$, the ratio $r_{pp}=S^2T_0G/K$ becomes the relevant parameter when the efficiency is maximized according to the voltage. Indeed, in that case the maximum of efficiency reads as $\eta_\mathrm{max}=\eta_C(1-\sqrt{1-r_{pp}})/(1+\sqrt{1-r_{pp}})$, where $\eta_C$ is the Carnot efficiency.
%Notice that in place of $ZT_0=S^2T_0G/\tilde K$, which is the relevant parameter when the efficiency is maximized according to the charge current in $\eta_{max}=\eta_C(\sqrt{1+ZT_0}-1)/(\sqrt{1+ZT_0}+1)$, the ratio $r_{pp}=S^2T_0G/K$ is the significant parameter when the efficiency is maximized according to the voltage. Indeed, the maximum of efficiency reads in that case as: $\eta_{max}=\eta_C(1-\sqrt{1-r_{pp}})/(1+\sqrt{1-r_{pp}})$, where $\eta_C$ is the Carnot efficiency.

%%%%%%%%%%%%%%%%%%%%%%%%%%
\subsection{High voltage regime}

In this regime, the first contribution in Eqs.~(\ref{charge_noise}) to (\ref{mixed_noise_2}), and the first and second contributions in Eq.~(\ref{heat_noise}) are negligible and we set $T_{L,R}$ to zero, thus:
\begin{widetext}
\begin{eqnarray}\label{See_zeroT}
&&\mathcal{S}^{ee}_{LL}=\frac{e^2}{h}\mathrm{sign}(V)\int^{eV/2}_{-eV/2}\mathcal{T}(\epsilon)[1-\mathcal{T}(\epsilon)]d\epsilon=-\mathcal{S}^{ee}_{LR}~,
\end{eqnarray}
\begin{eqnarray}\label{Seh_zeroT}
&&\mathcal{S}^{eh}_{LL}=\frac{e}{h}\mathrm{sign}(V)\int^{eV/2}_{-eV/2}\left(\epsilon-\frac{eV}{2}\right)\mathcal{T}(\epsilon)[1-\mathcal{T}(\epsilon)]d\epsilon=-\mathcal{S}^{he}_{LR}~,
\end{eqnarray}
\begin{eqnarray}\label{She_zeroT}
&&\mathcal{S}^{he}_{RR}=\frac{e}{h}\mathrm{sign}(V)\int^{eV/2}_{-eV/2}\left(\epsilon+\frac{eV}{2}\right)\mathcal{T}(\epsilon)[1-\mathcal{T}(\epsilon)]d\epsilon=-\mathcal{S}^{eh}_{LR}~,
\end{eqnarray}
\begin{eqnarray}\label{ShhLL_zeroT}
&&\mathcal{S}^{hh}_{LL}=\frac{1}{h}\mathrm{sign}(V)\int^{eV/2}_{-eV/2}\left(\epsilon-\frac{eV}{2}\right)^2\mathcal{T}(\epsilon)[1-\mathcal{T}(\epsilon)]d\epsilon~,
\end{eqnarray}
\begin{eqnarray}\label{ShhRR_zeroT}
&&\mathcal{S}^{hh}_{RR}=\frac{1}{h}\mathrm{sign}(V)\int^{eV/2}_{-eV/2}\left(\epsilon+\frac{eV}{2}\right)^2\mathcal{T}(\epsilon)[1-\mathcal{T}(\epsilon)]d\epsilon~,
\end{eqnarray}
\begin{eqnarray}\label{Shhcross_zeroT}
&&\mathcal{S}^{hh}_{LR}=-\frac{1}{h}\mathrm{sign}(V)\int^{eV/2}_{-eV/2}\left(\epsilon^2-\frac{e^2V^2}{4}\right)\mathcal{T}(\epsilon)[1-\mathcal{T}(\epsilon)]d\epsilon~.
\end{eqnarray}
\end{widetext}
In contrast to what happens in the linear response regime, the heat auto- and cross-correlators take distinct values. As a consequence, the auto-ratio $r_{pp}$ and the cross-ratio $r_{p\ne q}$ will differ. Thus, distinct values of $r_{pp}$ and $r_{p\ne q}$ is a signature that the system operates outside the linear response regime.

We now focus on two concrete nanosystems namely a quantum point contact and a quantum dot to further examine the correlators and the ratios of noises we have introduced and to interpret their values.

%%%%%%%%%%%%%%%%%%%%%%%%%%%%%%%%%%%%%%%%%%%%%%%%%%%%%%%%%%%%%%%%%%
%																 %
%																 %
%		QUANTUM POINT CONTACT									 %
%																 %
%																 %
%%%%%%%%%%%%%%%%%%%%%%%%%%%%%%%%%%%%%%%%%%%%%%%%%%%%%%%%%%%%%%%%%%

\section{Application to a quantum point contact}

\begin{figure}[h]
\begin{center}
\includegraphics[width=6cm]{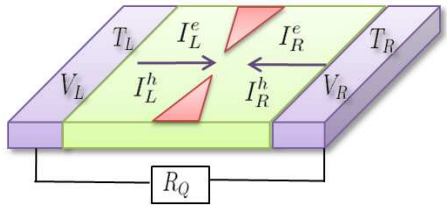}
\caption{Picture of the QPC coupled with an ohmic environment, and conventions chosen for left/right currents.}
\label{figure_QPC}
\end{center} 
\end{figure}

The first application of our results concerns a quantum point contact in an ohmic environment with resistance equal to $R_Q=h/e^2$ (see Fig.~\ref{figure_QPC}). The coupling to the ohmic environment leads to a drop in the conductance of the QPC at low voltage and temperature known as the dynamical Coulomb blockage. First predicted and experimentally verified for tunnel junctions \cite{devoret90,holst94}, it is present in the QPC \cite{levyyeyati01,pierre11}.
Because of this coupling, measured by $\Gamma$, its transmission coefficient acquires an energy dependency: $\mathcal{T}(\epsilon)=\epsilon^2/(\epsilon^2+\Gamma^2)$. Formally, this energy dependency can be obtained by the means of a mapping between such a system and a Luttinger liquid with a single impurity and interactions parameter equal to one-half \cite{kinderman03,safi04,zamoum12} allowing us to perform a refermionization procedure \cite{chamon96,vondelft98}. 
Since this system exhibits an electron-hole symmetry, the thermoelectric differential conductances $X$ and $Y$ are equal to zero \cite{note2}. We will show that it is the case for the mixed noises $\Sehpq$ and $\Shepq$ in the  linear response regime but not in the high voltage regime.

%%%%%%%%%%%%%%%%%%%%%%%%%%%%%
\subsection{Linear response regime}

We first focus on the case where the temperatures of the reservoirs are identical, $T_{L,R}=T_0$, and large in comparison to the applied voltage. Since $X$ and $Y$ are equal to zero, but not $G$ and $K$ (see Tab.~\ref{table1}), the resulting figure of merit such as the ratio of noises cancels because of Eq.~(\ref{ZT0}). From Eqs.~(\ref{See_linear}) to (\ref{Shh_linear}), we deduce the noises and give their equivalent expressions in Tab.~\ref{table1}. When the temperature is the largest energy scale of the problem, the electrical and thermal conductances take constant values: $G_Q$ and $K_Q$, where  $G_Q=e^2/h$ is the quantum of electrical conductance, and $K_Q=\pi^2k_B^2 T_0/3h$ is the quantum of thermal conductance recently measured in such a system \cite{pierre13}.

\begin{table}[!h]
\begin{center}
\begin{tabular}{|c||c|c|}
\hline
QPC& $\{eV,\Gamma\}\ll k_BT_0$ & $eV\ll k_BT_0\ll \Gamma $\\ \hline\hline
$G$& $G_Q$& $\frac{\pi^2}{3}\left(\frac{k_BT_0}{\Gamma}\right)^2G_Q$  \\ \hline
$X,Y$&0&0\\\hline
$K$ & $K_Q$& $\frac{7\pi^2}{5}\left(\frac{k_BT_0}{\Gamma}\right)^2K_Q$\\ \hline
$\begin{array}{ll}
\SeeLL  & =  -\SeeLR   \\
 &   = 2k_BT_0G
\end{array}$& $2k_BT_0G_Q$&  $\frac{2\pi^2}{3}\frac{(k_BT_0)^3}{\Gamma^2}G_Q$ \\ \hline
$\begin{array}{ll}
\SehLL & =-\SheLR \\
 &   = 2k_BT_0Y
\end{array}$ & $0$& $0$\\ \hline
$\begin{array}{ll}
\ShhLL & =-\ShhLR \\
 &   = 2k_BT_0^2K
\end{array}$ & $2k_BT_0^2K_Q$&$\frac{14\pi^2}{5}\frac{k_B^3T_0^4}{\Gamma^2}K_Q$\\ \hline
$r_{LL}=r_{RR}=r_{LR}$& 0 & 0 \\ \hline
\end{tabular}
\caption{QPC in the linear response regime -- Equivalent expressions of the differential conductances, 
noises
and ratios of noises obtained for $V=0$ and $T_{L,R}=T_0$.}
\label{table1}
\end{center}
\end{table}

\begin{figure}[h]
\begin{center}
\includegraphics[width=4.2cm]{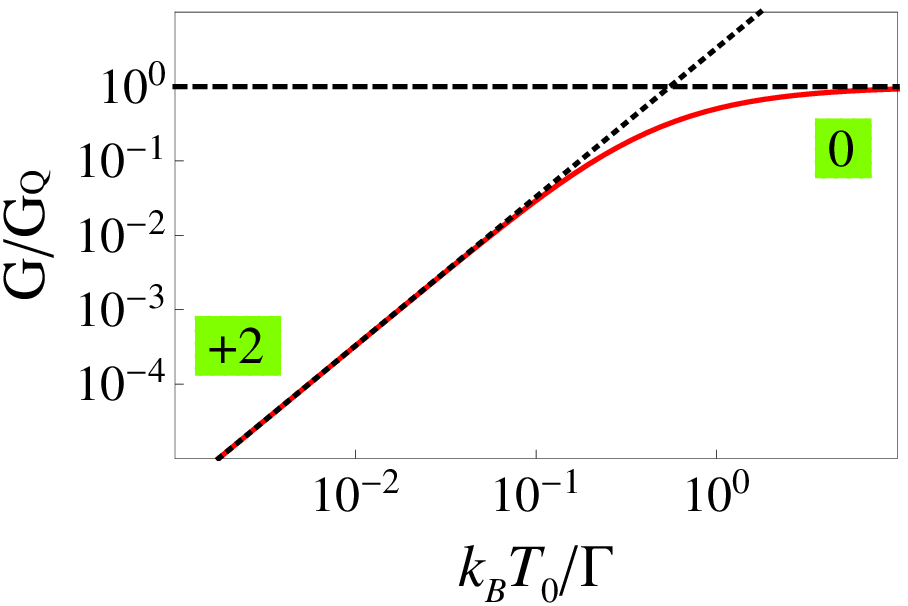}
\includegraphics[width=4.2cm]{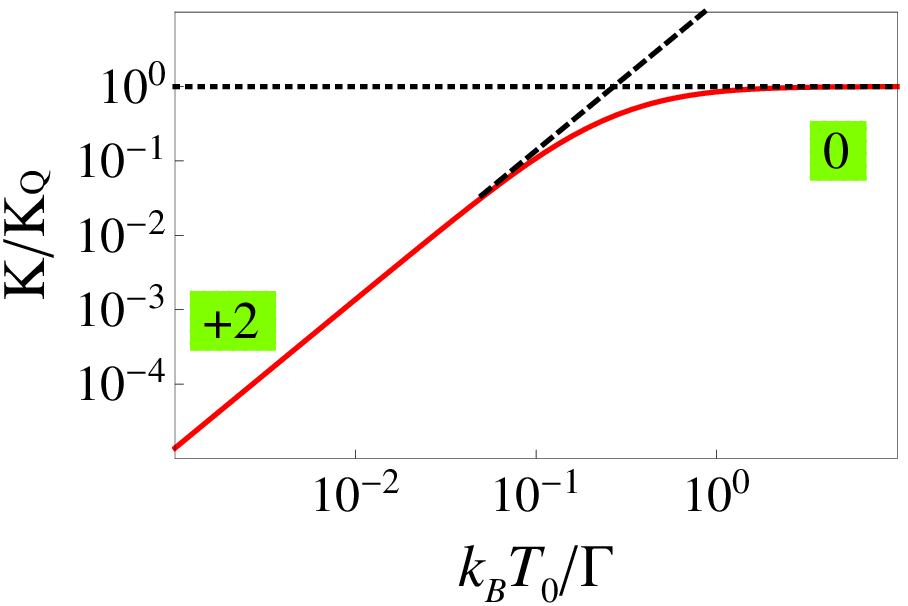}
\caption{QPC in the linear response regime -- Differential conductances $G$ and $K$ as a function of the average temperature $T_0$ for $V=0$ and $T=0$. $X$ and $Y$ are not shown since they are equal to zero in that regime. As a consequence, the associated ratios $r_{pq}$ cancel. The exponents of the power laws obtained in the equivalent expressions of Tab.~\ref{table1} are indicated in the green squares, and their signatures are represented by dashed lines.}
\label{figure_QPC_G_1}
\end{center} 
\end{figure}

Figure~\ref{figure_QPC_G_1} shows the crossover between the temperature power laws of the differential conductances $G$ and $K$ at strong coupling $\Gamma$ with the environment and their constant limits $G_Q$ and $K_Q$ at weak $\Gamma$ which corresponds to a QPC decoupled from the ohmic environment. Note that the Wiedemann-Franz relation between electrical and thermal conductances does not applied except when the temperature is the largest energy scale (see the central column of Tab.~ \ref{table1}). In that latter case:
\begin{eqnarray}
\frac{K}{GT_0}=\frac{K_Q}{G_QT_0}=\frac{\pi^2k_B^2}{3e^2}={\cal L}~,
\end{eqnarray}
which is the Lorenz factor. Identically, $\Shhpq/(\Seepq T_0)={\cal L}$ in that regime.

%%%%%%%%%%%%%%%%%%%%%%%%%%
\subsection{High voltage regime}

We turn now our interest to the case where the applied voltage is large in comparison to the temperature. In this limit, the integrals of Eqs.~(\ref{See_zeroT}) to (\ref{Shhcross_zeroT}) can be performed analytically (see Appendix A for the expressions of the currents and noises). The equivalent expressions of the differential conductances, noises and ratios of noises are given in Tab.~\ref{table2}. 
Again, only the electrical and thermal conductances are relevant in this regime since $X$ and $Y$ are still zero due to  the electron-hole symmetry. We see that $G$ does not depend on voltage when $eV$ is the largest energy of the problem (central column of Tab.~\ref{table2}).

\begin{figure}[h]
\begin{center}
\includegraphics[width=4.4cm]{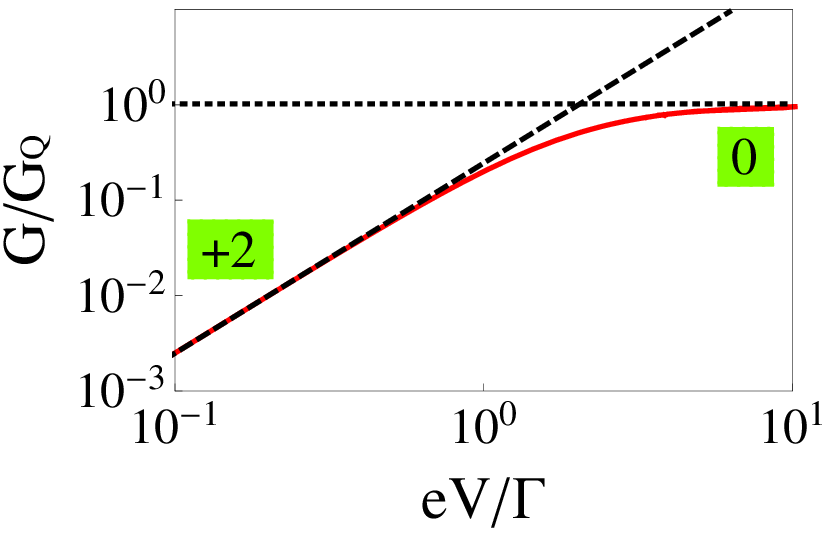}
\includegraphics[width=4cm]{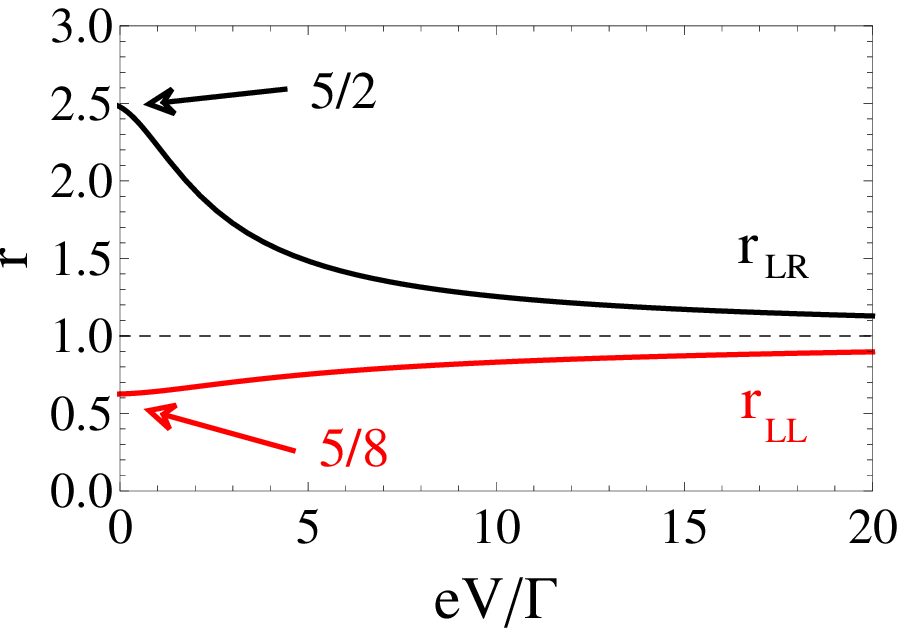}
\caption{QPC in the high voltage regime -- Left: differential conductance $G$ (red line), power law exponents (in green squares) and signatures (dashed lines) of their equivalent expressions given Tab.~\ref{table2}. Right: noises ratios ($r_{LL}=r_{RR}$) as a function of the voltage at zero temperatures.}
\label{figure_QPC_G_2}
\end{center} 
\end{figure}

Due to the parity in energy of the QPC transmission $\mathcal{T}$, the heat auto-correlators do not depend on the reservoir: $\ShhLL=\ShhRR$. This property ensures identical auto-ratios $\rLL=\rRR$. Comparing the central and the last columns of Tab.~\ref{table2}, we note that $\SehLL= -(V/2)\SeeLL$ and $\ShhLL\propto -(V/2)\SehLL$ (idem for the cross-noises).  The proportionality coefficient is exactly $1$ when $\{k_BT_0,\Gamma\}\ll eV$, while for $k_BT_0\ll eV\ll \Gamma$, it is above one ($8/5$) for the auto-noises and below one ($2/5$) for the cross-noises, which gives $\rLL<1$ and $\rLR>1$ respectively (see the central and the last columns of Tab.~\ref{table4}). As already mentioned in Sec. II, the fact that $\rpq=1$ means that the second contribution in Eq.~(\ref{energy}) cancels. Since we have $\mathcal{S}^{eE}_{pq}=\mathcal{S}^{Ee}_{pq}=0$ for a QPC due to electron-hole symmetry, it leads to $\mathcal{S}^{EE}_{pq}=0$ in full agreement with the fact that in a QPC decoupled from its environment ($\Gamma\rightarrow 0$) at zero temperature ($T\rightarrow 0$), there is no mechanism, neither thermal excitations nor coupling to environment, that allows energy to fluctuate. In contrast, when the coupling to the environment increases, the value of the ratios moves away from one due to the appearance of energy fluctuations (see the right panel of Fig.~\ref{figure_QPC_G_2}).

\begin{table}[!h]
\begin{center}
\begin{tabular}{|c||c|c|}
\hline
QPC& $\{k_BT_0,\Gamma\}\ll eV$ & $k_BT_0\ll eV\ll \Gamma $\\ \hline\hline
$G$& $G_Q$& $\left(\frac{eV}{2\Gamma}\right)^2G_Q$  \\ \hline
$X,Y$& $0$ & $0$\\ \hline
$K$& $K_Q$ & $\left(\frac{eV}{2\Gamma}\right)^2K_Q$\\ \hline
$\SeeLL =-\SeeLR$& $\frac{\pi\Gamma}{2}G_Q$& $\frac{e^3|V|^3}{12\Gamma^2} G_Q=\frac{e|V|}{3} G$ \\ \hline
$\begin{array}{ll}
\SehLL & =-\SehRR \\
 &  =-\SheLR
\end{array}$ &$-\frac{\pi\Gamma V}{4}G_Q=-\frac{V}{2}\SeeLL$& $-\frac{e^3|V|^3V}{24\Gamma^2} G_Q=-\frac{V}{2}\SeeLL$\\ \hline
$\ShhLL=\ShhRR$ & $\frac{\pi\Gamma V^2}{8}G_Q=\frac{V^2}{4}\SeeLL$ & $\frac{e^3|V|^5}{30\Gamma^2} G_Q= \frac{2V^2}{5}\SeeLL$\\ \hline
$\ShhLR$ & $\frac{\pi\Gamma V^2}{8}G_Q=-\frac{V^2}{4}\SeeLR$ & $\frac{e^3|V|^5}{120\Gamma^2} G_Q= -\frac{V^2}{10}\SeeLR $\\ \hline
 $\rLL=\rRR$  & 1 & 5/8 \\ \hline
 $\rLR$ & 1 & 5/2 \\ \hline
\end{tabular}
\caption{QPC in the high voltage regime -- Equivalent expressions of the differential conductances, noises and ratios of noises obtained for $T_{L,R}$ going to zero.}
\label{table2}
\end{center}
\end{table}

The left graph of Fig.~\ref{figure_QPC_G_2} shows the variation of the electrical conductance as a function of the voltage. By comparing the last columns of Tabs.~\ref{table1} and \ref{table2}, we note that the power law exponent obtained in the limit $k_BT_0\ll eV\ll\Gamma$, for which $G\propto (eV/\Gamma)^2$, is the same as the one obtained in the limit $eV\ll k_BT_0\ll\Gamma$ for which $G\propto (k_BT_0/\Gamma)^2$, meaning that temperature and voltage play a similar role for the electrical conductance \cite{kane92}. The same occurs for the thermal conductance.

In contrast to what happens in the linear response regime, the ratios of noises are non-zero in the high voltage regime. Interestingly, whereas $r_{LL}$ stays below one when varying the voltage, the cross-ratio $r_{LR}$ exhibits a value larger than one (up to $5/2$) in the strong coupling limit as shown in the right graph of Fig.~\ref{figure_QPC_G_2}, where the equivalent expressions given in Tab.~\ref{table2} are recovered in both $eV/\Gamma\ll 1$ and $eV/\Gamma\gg 1$ limits.  It confirms the fact that auto-ratio and cross-ratio differ in the non-linear regime, as explained in Sec. IV.B.

%%%%%%%%%%%%%%%%%%%%%%%%%%%%%%%%%%%%%%%%%%%%%%%%%%%%%%%%%%%%%%%%%%
%																 %
%																 %
%		QUANTUM DOT										    	 %
%																 %
%																 %
%%%%%%%%%%%%%%%%%%%%%%%%%%%%%%%%%%%%%%%%%%%%%%%%%%%%%%%%%%%%%%%%%%

\section{Application to a quantum dot}

\begin{figure}[h]
\begin{center}
\includegraphics[width=6cm]{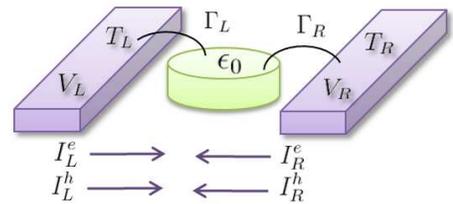}
\caption{Picture of the QD and conventions chosen for left/right currents.}
\label{figure_QD}
\end{center} 
\end{figure}

We now consider a single level non-interacting quantum dot with a transmission coefficient $\mathcal{T}(\epsilon)=\Gamma^2/[(\epsilon-\epsilon_0)^2+\Gamma^2]$, where $\epsilon_0$ is the energy level of the dot (see Fig.~\ref{figure_QD}), and $\Gamma$ is the broadening due to the contact to the reservoirs which is assumed to be energy independent and symmetrical $\Gamma_L=\Gamma_R=\Gamma$.

In the following subsections, we study the behavior of the QD in various regimes: in the linear response regime (small $T$ and $V$), in the high voltage regime (small $T$ and $T_0$), then in the Schottky regime which corresponds to a dot weakly coupled to the reservoirs (small $\Gamma$), and finally, in the intermediate regime when all the characteristic energies of the system are of the same order of magnitude.

%%%%%%%%%%%%%%%%%%%%%%%%%%%%
\subsection{Linear response regime}

\begin{figure}[h]
\begin{center}
\includegraphics[width=4.2cm]{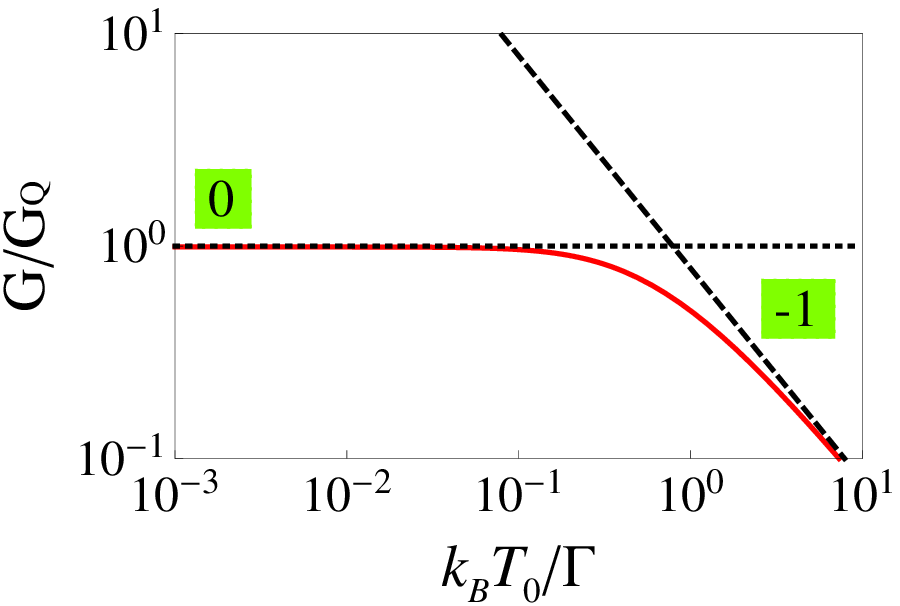}
\includegraphics[width=4.2cm]{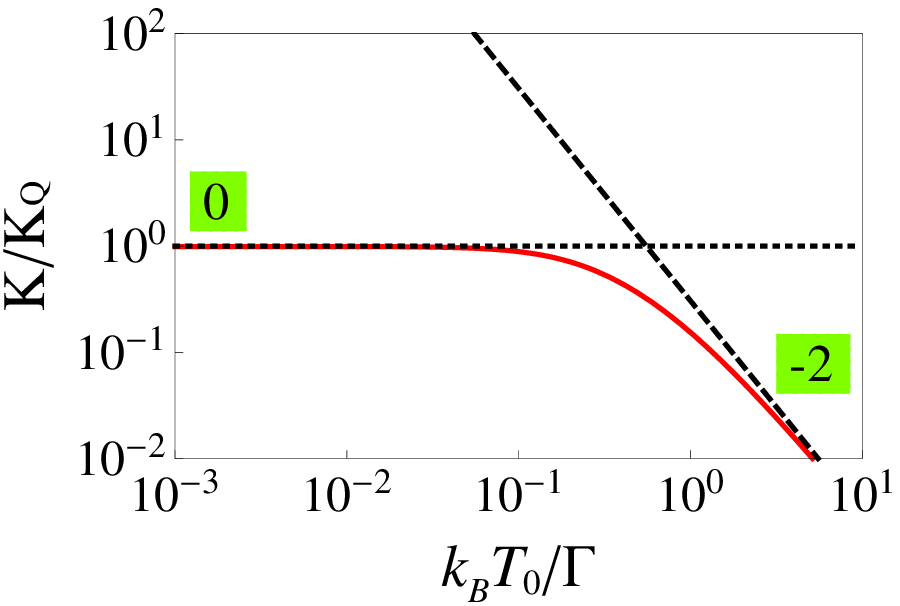}
\includegraphics[width=4.2cm]{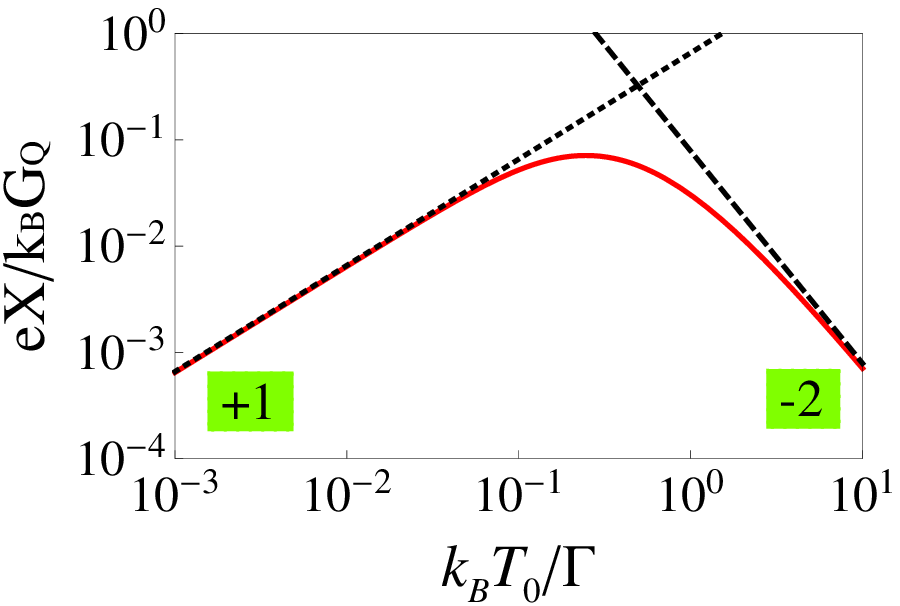}
\includegraphics[width=4.2cm]{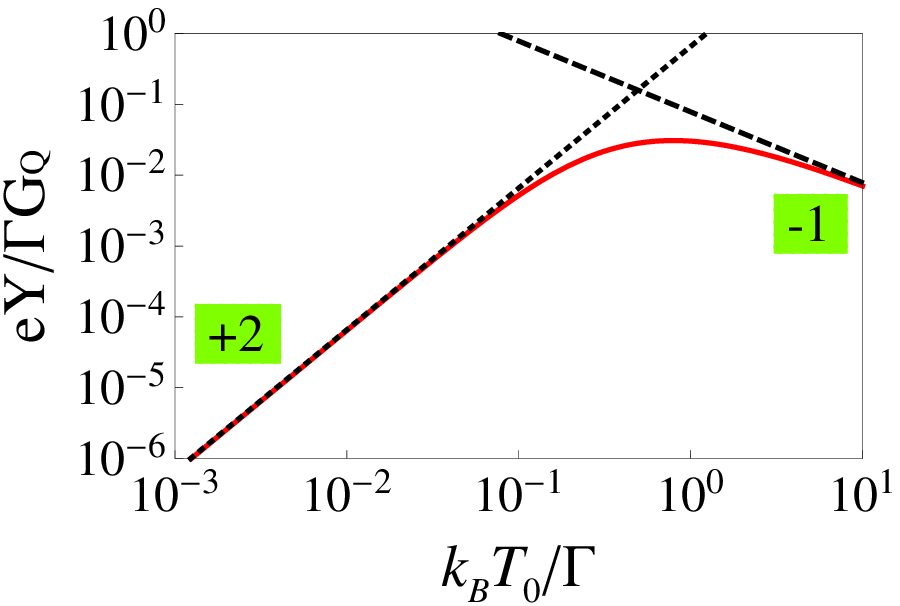}
\caption{QD in the linear response regime -- Differential conductances as a function of the average temperature for $V=0$, $T=0$, and $\epsilon_0/\Gamma=0.1$. Power law exponents (in green squares) and signatures (dashed lines) of their equivalent expressions given in Tab.~\ref{table3}.}
\label{figure_QD_G}
\end{center} 
\end{figure}

In Fig.~\ref{figure_QD_G} are plotted differential conductances as well as their equivalent expressions summarized in Table~\ref{table3}. For both limits $k_BT_0\ll \Gamma$ and $k_BT_0\gg\Gamma$, these quantities exhibit power laws with various exponents. Note that $X$, $Y$, $S^{eh}_{pq}$, $S^{he}_{pq}$  and $r_{pq}$ all cancel when electron-hole symmetry applies, i.e. $\epsilon_0=0$, and thus we recover for the noises the results obtained in the QPC since in that case $\mathcal{T}_{\mathrm{QD}}(\epsilon)=1-\mathcal{T}_{\mathrm{QPC}}(\epsilon)$ (compare the central column of Tab.~\ref{table1} and the last column of Tab.~\ref{table3}).

\begin{table}[!h]
\begin{center}
\begin{tabular}{|c||c|c|}
\hline
QD& $\{eV,\Gamma\}\ll k_BT_0$ & $eV\ll k_BT_0\ll \Gamma $\\ \hline\hline
$G$& $\frac{\pi}{4}\left(\frac{\Gamma}{k_BT_0}\right)G_Q$ & $G_Q$ \\ \hline
$X=Y/T_0$& $\frac{\pi k_B }{4e} \left(\frac{\epsilon_0\Gamma}{k_B^2T_0^2}\right)G_Q$ & $\frac{2e\epsilon_0}{\Gamma^2} K_Q$\\ \hline
$K$& $\frac{3}{\pi^2}\left(\frac{\Gamma}{k_BT_0}\right)^2K_Q$ & $K_Q$\\ \hline
$\begin{array}{ll}
\SeeLL  & =  -\SeeLR   \\
 &   = 2k_BT_0G
\end{array}$& $\frac{\pi \Gamma}{2} G_Q$ & $2k_BT_0G_Q$\\ \hline
$\begin{array}{ll}
\SehLL & =-\SheLR \\
 &   = 2k_BT_0Y
\end{array}$& $\frac{\pi\Gamma \epsilon_0}{2e}  G_Q$ & $\frac{4\pi^2}{3e}\left(\frac{k_BT_0}{\Gamma}\right)^3\Gamma\epsilon_0G_Q$\\ \hline
$\begin{array}{ll}
\ShhLL & =-\ShhLR \\
 &   = 2k_BT_0^2K
\end{array}$& $\frac{6\Gamma^2}{\pi^2k_B}K_Q$ & $2k_BT_0^2K_Q$\\ \hline
$r_{LL}=r_{RR}=r_{LR}$& $\frac{\pi}{4}\frac{\epsilon_0^2}{\Gamma k_BT_0}=\frac{XY}{GK}$ & $\frac{4\pi^2}{3}\left(\frac{k_BT_0\epsilon_0}{\Gamma^2}\right)^2=\frac{XY}{GK}$ \\ \hline
\end{tabular}
\caption{QD in the linear response regime -- Equivalent expressions of the differential conductances, noises and ratios of noises obtained for $V=0$, $T_{L,R}=T_0$, and $\varepsilon_0\ll\Gamma$.}
\label{table3}
\end{center}
\end{table}

\begin{figure}[!h]
\begin{center}
\includegraphics[width=6cm]{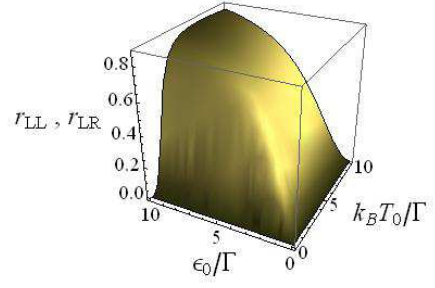}
\caption{QD in the linear response regime -- Variation of the ratios $\rLL=r_{RR}=\rLR$ as a function of temperature $T_0$ and dot energy level $\epsilon_0$ for equal left and right temperatures at zero voltage.}
\label{figure_QD_ratio1_3D}
\end{center} 
\end{figure}

From Tab.~\ref{table3}, we see that the differential conductances $X$ and $Y$ are proportionals to the dot energy level in agreement with the fact that thermoelectric measurements in the high temperature regime give indication on the position of the dot energy level relative to the Fermi energy \cite{paulsson03}. We find here that because of the fluctuation-dissipation theorem, the mixed noises are themselves proportional to the dot energy level. In addition, the ratios $\rLL$ and $\rLR$ are equal to each other and they correspond to $XY/GK$. Figure~\ref{figure_QD_ratio1_3D} shows the evolution of these ratios as a function of $T_0$ and $\epsilon_0$ in the absence of any temperature gradient. They vanish at $\epsilon_0=0$ as expected and their maximum does not exceed one even for large values of $\epsilon_0$ because of the Cauchy-Schwarz inequality.

%%%%%%%%%%%%%%%%%%%%%%%%%%
\subsection{High voltage regime}

In that regime, the differential conductances read as:
\begin{eqnarray}
G&=&\frac{G_Q}{2}\sum_{\pm}\mathcal{T}\left(\pm\frac{eV}{2}\right)~,\\
X&=&-\frac{eK_Q}{\Gamma^2}\sum_{\pm}\left(\pm\frac{eV}{2}-\epsilon_0\right)\mathcal{T}^2\left(\pm\frac{eV}{2}\right)~,\\
K&=&\frac{K_Q}{2}\sum_{\pm}\mathcal{T}\left(\pm\frac{eV}{2}\right)~.
\end{eqnarray}
The integrals of Eqs.~(\ref{See_zeroT}) to (\ref{Shhcross_zeroT}) giving the noises are performed analytically (see Appendix B for the expressions of the currents and noises). 
The equivalent expressions of conductances and noises are given in Tab.~\ref{table4}. 
The same as in the QPC, we find that heat, charge and mixed noises are strongly related to each other via the voltage, since we have $\SehLL= -(V/2)\SeeLL$  and $\SheLR= -(V/2)\SeeLR$ in all cases,  and moreover
$\ShhLL\propto -(V/2)\SehLL\propto (V^2/4)\SeeLL$ and $\ShhLR\propto (V/2)\SheLR\propto -(V^2/4)\SeeLR$. For a strict equality, both auto- and cross-ratios of noises reach one. Otherwise, the proportionality coefficients give $\rLL<1$ and $\rLR>1$ and stay unchanged as long as $eV$ stays the lowest energy of the problem excluding temperature  (compare the second and third columns of Tab.~\ref{table4}).

\begin{figure}[!h]
\begin{center}
\includegraphics[width=4.25cm]{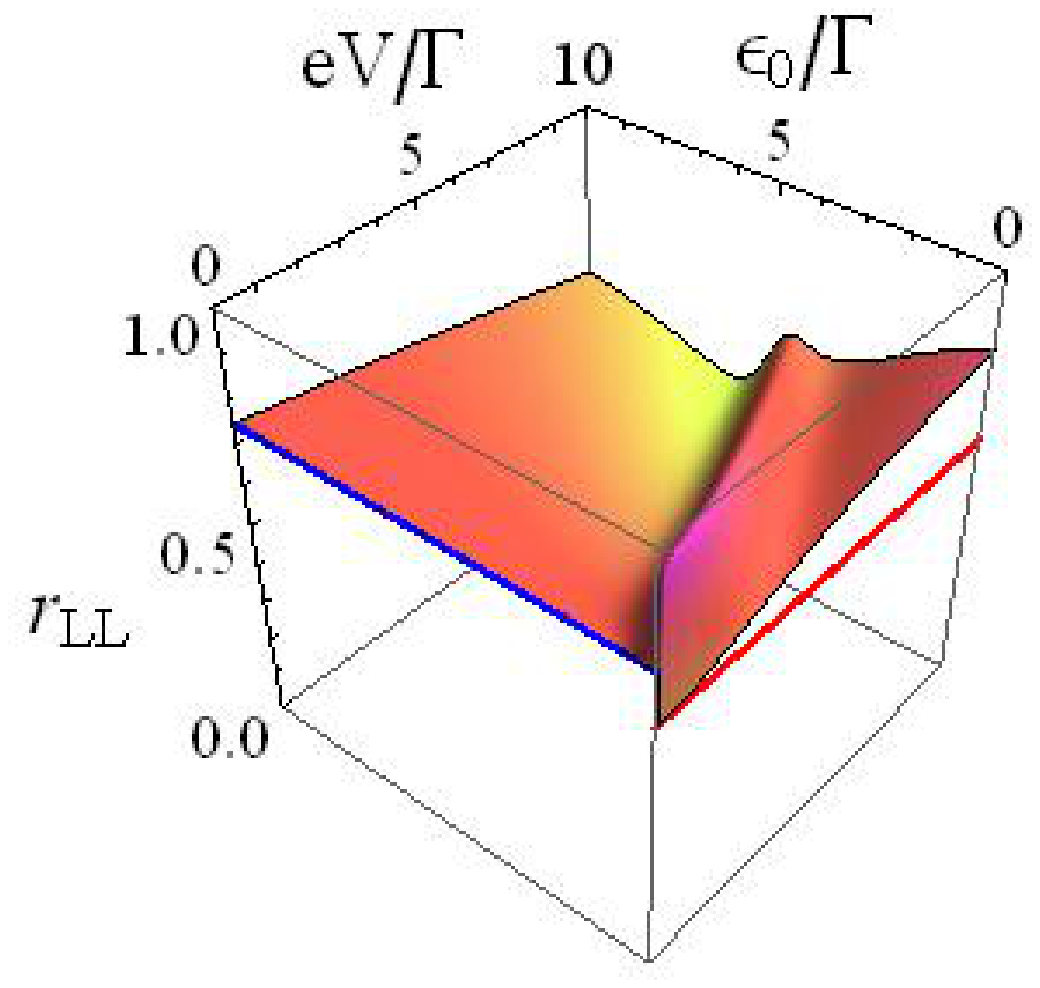}
\includegraphics[width=4.25cm]{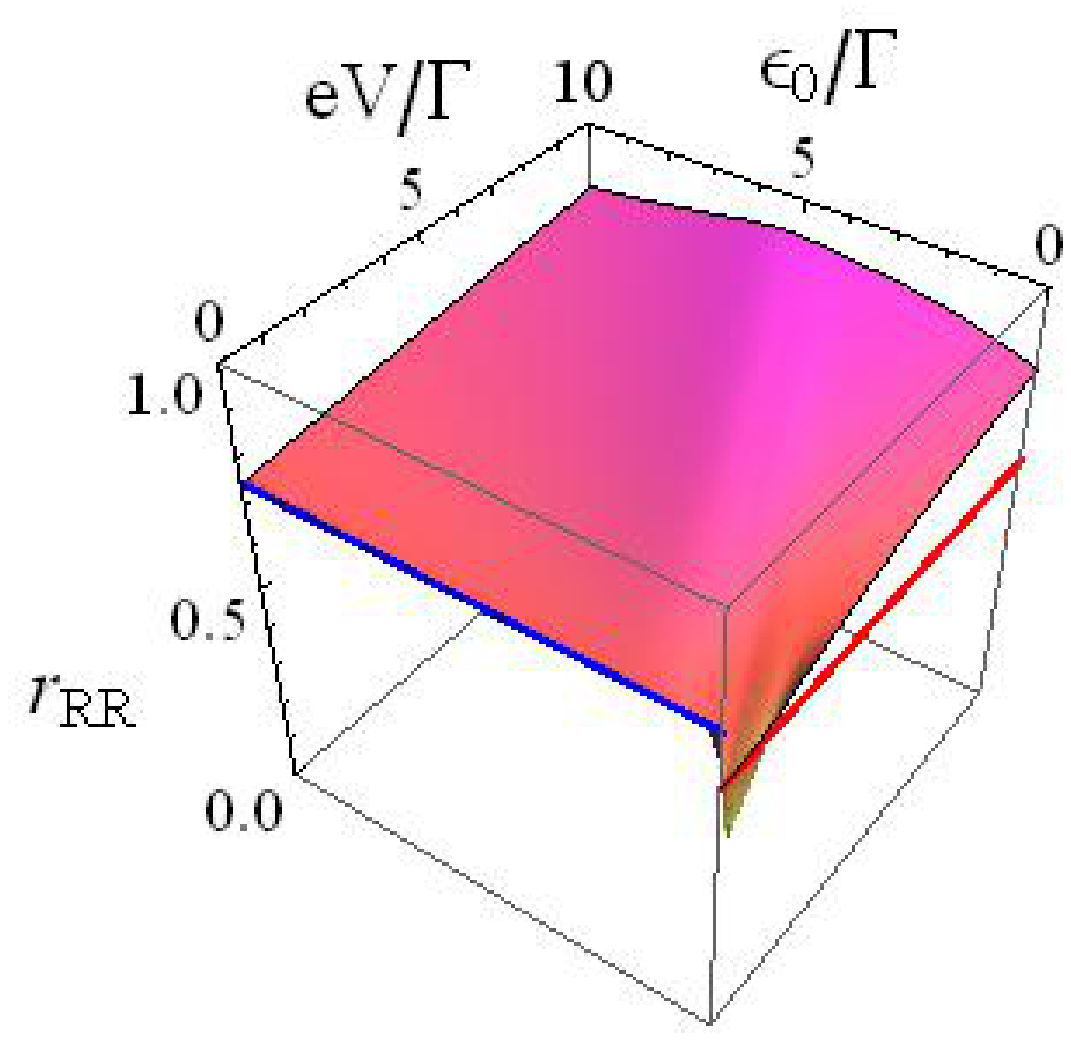}
\includegraphics[width=4.5cm]{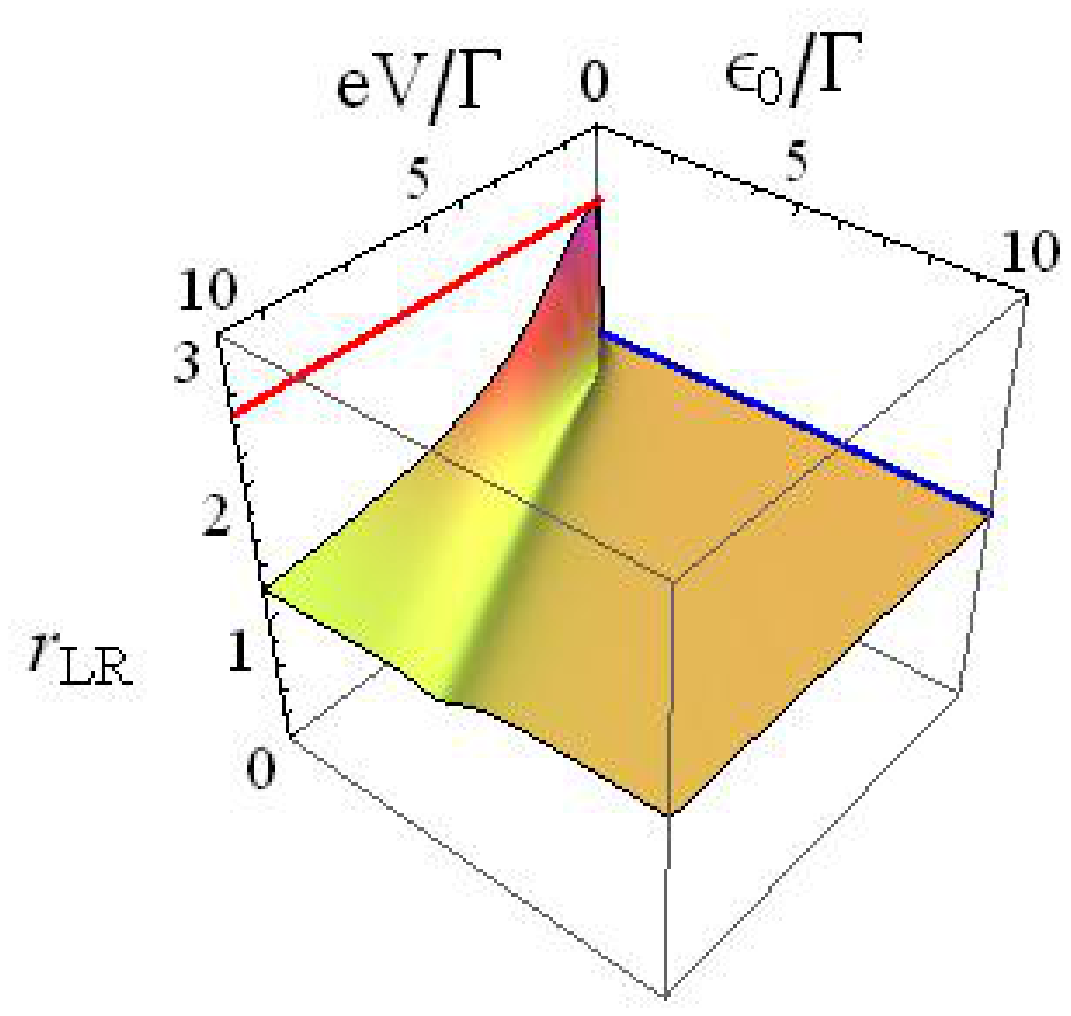}
\caption{QD in the high voltage regime -- Variation of the ratios $\rLL$, $\rRR$ and $\rLR$ as a function of voltage and dot energy level at zero temperature. On both graphs, the red lines indicates their values in the limit $\epsilon_0=eV=0$ (i.e., $\rLL=r_{RR}=5/8$ and $\rLR=5/2$) and the blue lines in the limit $eV\ll \epsilon_0$ (i.e., $\rLL=r_{RR}=3/4$ and $\rLR=3/2$).}
\label{figure_QD_ratio2_3D}
\end{center} 
\end{figure}

In Tab.~\ref{table4}, the expressions for $\ShhLL$ and $\ShhRR$, and hence for auto-ratios, are identical in the three limits we consider. This is no longer the case in intermediate regimes,  in contrast with the QPC, as shown Figs.~\ref{figure_QD_ratio2_3D} and  \ref{figure_QD_ratio3_3D}. Indeed, $\mathcal{T}_{QD}(\epsilon)$ is no longer an even function when $\epsilon_0 \neq 0$ which leads to $\ShhLL \neq \ShhRR$.

\begin{widetext}

\begin{table}[!h]
\begin{center}
\begin{tabular}{|c||c|c|c|c|}
\hline
QD & $\{eV , \Gamma\} \ll \epsilon_0 $ &   $ \{eV , \epsilon_0\} \ll \Gamma $ &   $ \{\Gamma, \epsilon_0\} \ll eV $ \\ \hline\hline
$G$& $\frac{\Gamma^2}{\epsilon_0^2}G_Q$ & $G_Q$ &$\frac{4\Gamma^2}{e^2V^2}G_Q$\\ \hline
$X=Y/T_0$& $\frac{2e\Gamma^2}{\epsilon_0^3}K_Q$&$\frac{2e\epsilon_0}{\Gamma^2}K_Q$ &$-\frac{24\epsilon_0\Gamma^2}{e^3V^4}K_Q$\\ \hline
$K$&$\frac{\Gamma^2}{\epsilon_0^2}K_Q$ &$K_Q$ &$\frac{4\Gamma^2}{e^2V^2}K_Q$\\ \hline
$\SeeLL =-\SeeLR$& $  (\frac{\Gamma}{\epsilon_0})^2 e|V|  G_Q=e|V|G$ & $  (\frac{\epsilon_0}{\Gamma})^2 e|V|  G_Q$  &$\frac{\pi\Gamma}{2}G_Q=\frac{\pi e^2V^2}{2\Gamma}G$\\ \hline
$\SehLL=-\SheLR$& $- (\frac{\Gamma}{\epsilon_0})^2 \frac{eV|V|}{2} G_Q=-\frac{V}{2}\SeeLL  $& $- (\frac{\epsilon_0}{\Gamma})^2 \frac{eV|V|}{2} G_Q=-\frac{V}{2}\SeeLL  $ & $-\frac{\pi\Gamma V}{4}G_Q=-\frac{V}{2}\SeeLL$\\ \hline
$\ShhLL=\ShhRR$& $ (\frac{\Gamma}{\epsilon_0})^2  \frac{e|V|^3}{3} G_Q=\frac{V^2}{3}\SeeLL  $  & $ (\frac{\epsilon_0}{\Gamma})^2  \frac{e|V|^3}{3} G_Q=\frac{V^2}{3}\SeeLL  $ & $\frac{\pi\Gamma V^2}{8}G_Q=\frac{V^2}{4}\SeeLL$\\ \hline
$\ShhLR$& $ (\frac{\Gamma}{\epsilon_0})^2  \frac{e|V|^3}{6} G_Q=-\frac{V^2}{6}\SeeLR  $  & $ (\frac{\epsilon_0}{\Gamma})^2  \frac{e|V|^3}{6} G_Q=-\frac{V^2}{6}\SeeLR  $  & $\frac{\pi\Gamma V^2}{8}G_Q=-\frac{V^2}{4}\SeeLR$\\ \hline
$\rLL=\rRR$& $3/4 $ & $3/4 $ &$1$\\ \hline
$\rLR$& $ 3/2 $ & $3/2 $ &$1$\\ \hline
\end{tabular}
\caption{High voltage regime in a QD -- Equivalent expressions of the differential conductances, noises and ratios of noises obtained for $T_{L,R}$ going to zero. We stress that heat auto-correlators and auto-ratios reach identical expressions only in the limits reported in this table (see Figs.~\ref{figure_QD_ratio2_3D} and \ref{figure_QD_ratio3_3D}).}
\label{table4}
\end{center}
\end{table}

\end{widetext}

Comparing auto- and cross-ratios of Fig.~\ref{figure_QD_ratio2_3D}, we see that they take distinct values in the high voltage regime, inversely to what happens in the high temperature regime, because of the distinct values taken by $\ShhLL$ and $\ShhLR$. The same as for the QPC, the cross-ratio $r_{LR}$ can have a value larger than one, whereas $r_{LL}$ and $\rRR$ stay always smaller than one, in agreement with the Cauchy-Schwarz inequality. At zero voltage and non-zero $\epsilon_0$, we recover the fractional values $3/4$ for $\rLL$ ($\rRR$), and $3/2$ for $\rLR$ as expected from Tab.~\ref{table4}. At both zero voltage and dot energy level, we recover the fractional values $5/8$ for $\rLL$ ($\rRR$), and $5/2$ for $\rLR$ as expected from Tab.~\ref{table2}.

%%%%%%%%%%%%%%%%%
\subsection{Schottky regime}

The Schottky regime is interesting to consider since in this case the noises are proportional to the currents. Indeed, in the limit of weak transmission, i.e., when $\Gamma$ is the smallest energy scale of the problem, and assuming a positive voltage in order to avoid the question of sign, we get:
\begin{eqnarray}
\label{schottky1}
\mathcal{S}^{ee}_{LL}&=&\mathcal{C}eI_L^e=-\mathcal{S}^{ee}_{LR}~,\\
\label{schottky2}
\mathcal{S}^{eh}_{LL}&=&\mathcal{C}eI_L^h=\mathcal{C}(\epsilon_0-\mu_{L})I^e_{L}=-\mathcal{S}^{eh}_{RL}~,\\
\label{schottky3}
\mathcal{S}^{eh}_{RR}&=&-\mathcal{C}eI_R^h=\mathcal{C}(\epsilon_0-\mu_{R})I^e_{L}=-\mathcal{S}^{eh}_{LR}~,\\
\label{schottky4}
\mathcal{S}^{hh}_{LL}&=&\mathcal{C}(\epsilon_0-\mu_L)I_L^h~,\\
\label{schottky5}
\mathcal{S}^{hh}_{LR}&=&-\mathcal{C}(\epsilon_0-\mu_L)I_R^h=\mathcal{S}^{hh}_{RL}~,
\end{eqnarray}
where $\mathcal{C}=\coth[(\epsilon_0-\mu_R)/2k_BT_R-(\epsilon_0-\mu_L)/2k_BT_L]$ is a thermal coefficient which reduces to one at zero temperature. Equations (\ref{schottky1}) to (\ref{schottky5}) lead to $r_{pq}=1$ meaning that we have a maximum of heat-charge correlation in the reservoirs. In addition, the heat and charge currents are themselves proportional: $I^h_p=(\epsilon_0-\mu_p)I^e_p/e$, in agreement with what is obtained in the tight charge/energy coupling (see for example Ref.~\cite{esposito09}). From these relations, it is possible to express the thermoelectric efficiency fully in terms of noises using the relation $eV=\mathcal{S}^{eh}_{LR}/I^e_{L}-\mathcal{S}^{hh}_{LR}/I^h_{R}$  derived from Eqs.~(\ref{schottky3}) and (\ref{schottky5}). For a refrigerator or heat pump working, the efficiency is defined as a the ratio of the output thermal power $P^\mathrm{th}_\mathrm{out}=|I^h_R|$ to the input electrical power $P^\mathrm{el}_\mathrm{in}=|VI^e_L|$. From $\eta=P^\mathrm{th}_\mathrm{out}/P^\mathrm{el}_\mathrm{in}$ we get:
\begin{eqnarray}\label{eta}
\eta=\frac{(\mathcal{S}^{eh}_{LR})^2}{\left|\mathcal{S}^{ee}_{LR}\mathcal{S}^{hh}_{LR}-(\mathcal{S}^{eh}_{LR})^2\right|}~.
\end{eqnarray}
This result is remarkable since it shows that even far from equilibrium, the thermoelectric efficiency is given by the noises. 
This expression differs from the one obtained in the linear response regime in two ways: the absence of the square roots  in the expression of the efficiency and the reservoirs indices. Indeed, in Eq.~(\ref{ZT0}) the indices play no role whereas the cross-noises appear in Eq.~(\ref{eta}) as they evidence the thermoelectric transfers from one reservoir to the other.
Moreover, the expression of Eq.~(\ref{eta}) suggests to introduce another cross-ratio $\tilde{r}_{pq}=(\Sehpq)^2/(\Seepq\Shhpq)$ instead of $\rpq=\Sehpq\Shepq/(\Seepq\Shhpq)$. Both are equal in the linear response regime.

%%%%%%%%%%%%%%%%%
\subsection{Intermediate regime}

\begin{figure}[!h]
\begin{center}
\includegraphics[width=4.2cm]{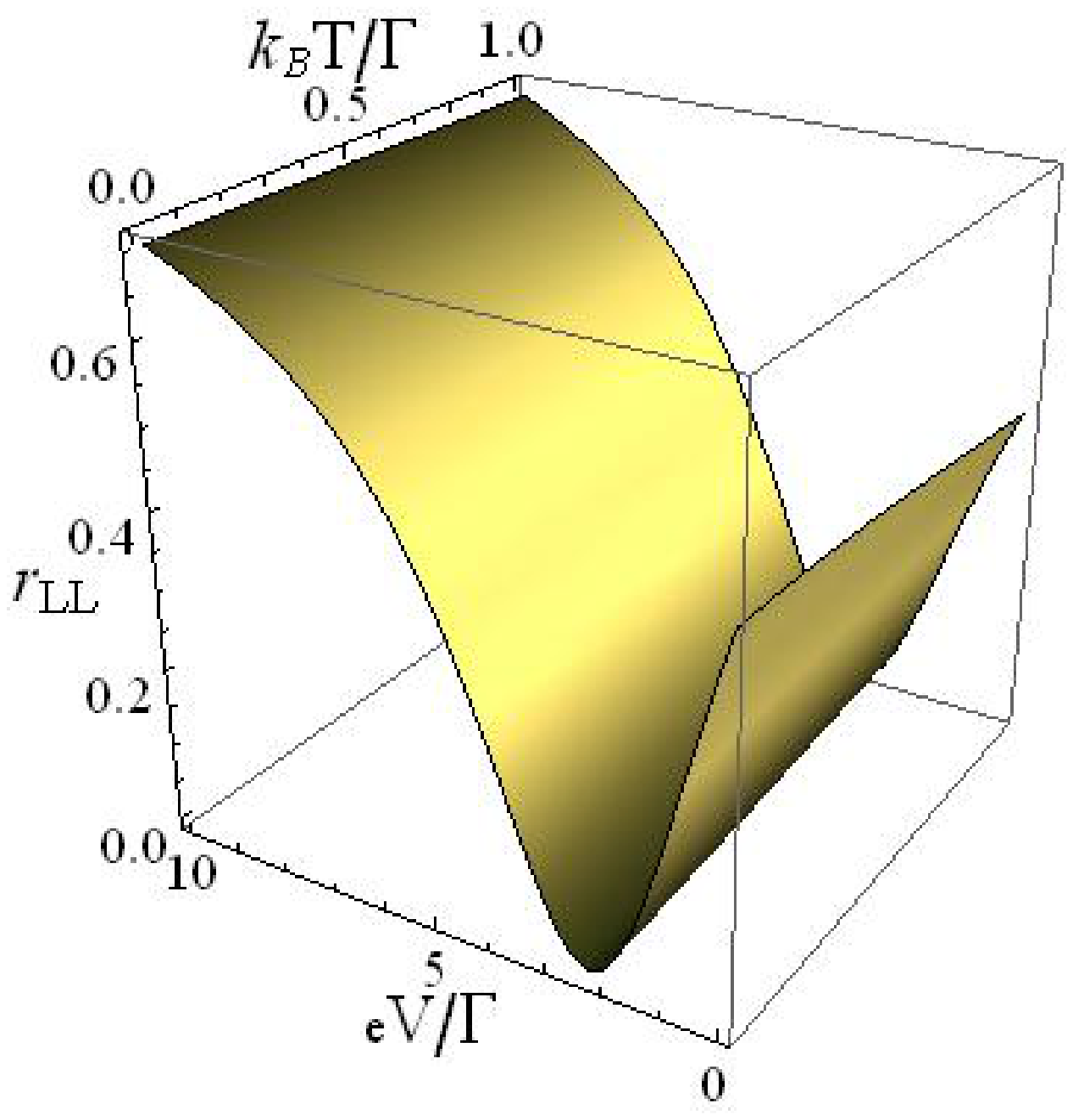}
\includegraphics[width=4.2cm]{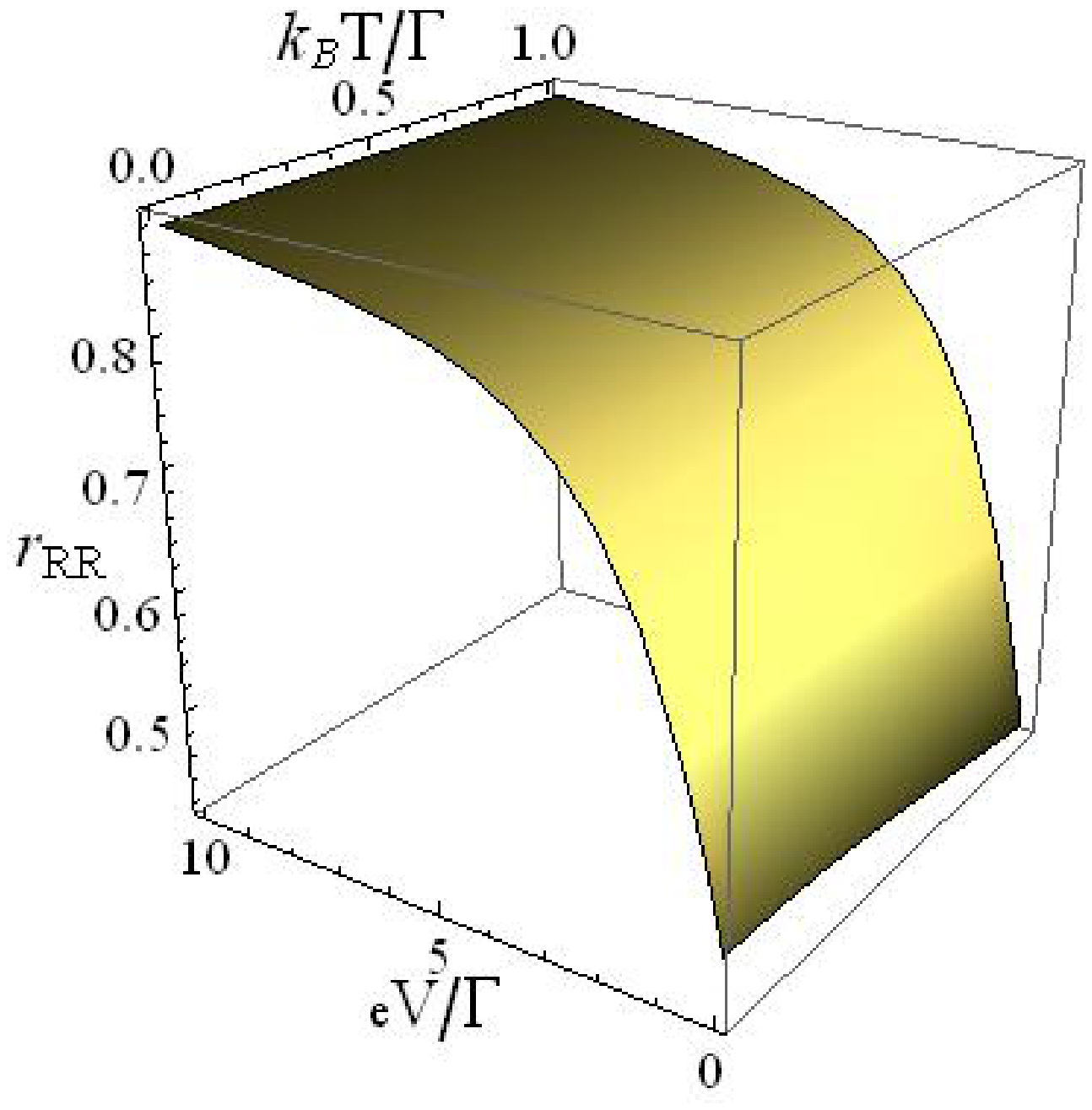}
\includegraphics[width=4.4cm]{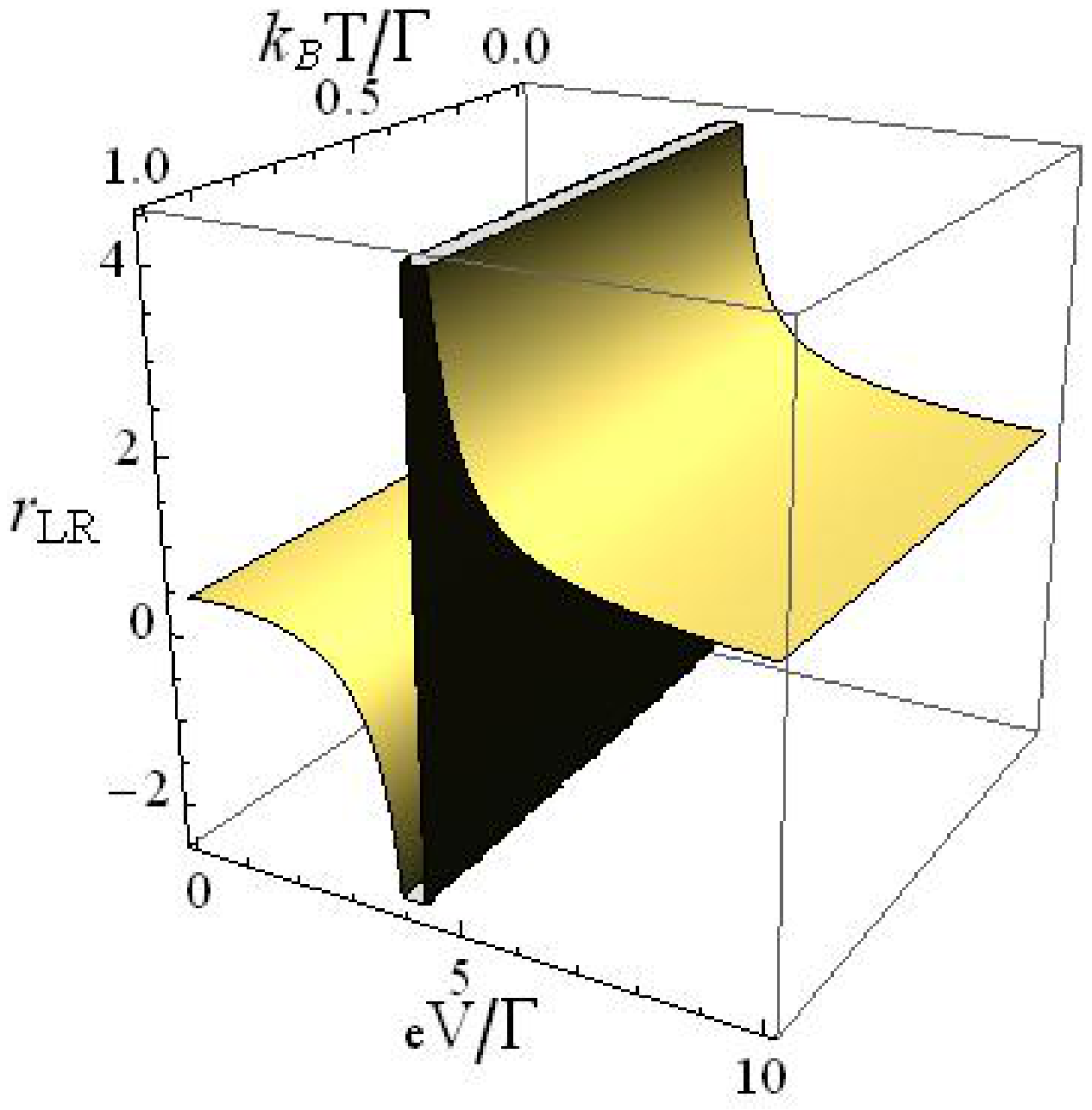}
\caption{QD in the intermediate regime -- Variation of the ratios $\rLL$, $\rRR$ and $\rLR$ as a function of the voltage and the temperature gradients. We take $\epsilon_0/\Gamma=2$ and $k_BT_0/\Gamma=1$.}
\label{figure_QD_ratio3_3D}
\end{center} 
\end{figure}

\begin{figure}[!h]
\begin{center}
\includegraphics[width=8.5cm]{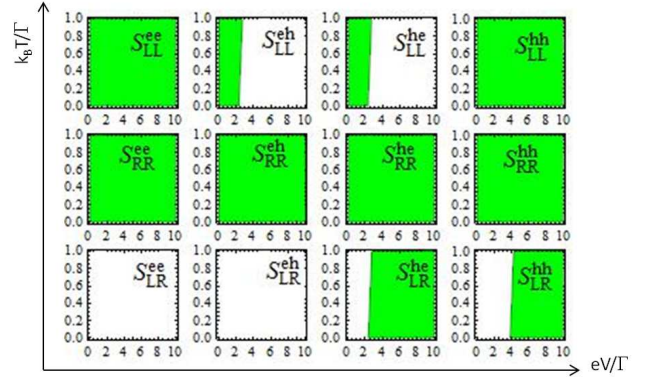}
\caption{QD in the intermediate regime -- Regions of positive sign (in green) for auto- and cross-correlators at $\epsilon_0/\Gamma=2$ and $k_BT_0/\Gamma=1$. Mixed noises $\mathcal{S}_{LL}^{eh}$, $\mathcal{S}_{LL}^{he}$ and $\mathcal{S}_{LR}^{he}$ change sign along the same line. The heat cross-noise $\mathcal{S}_{LR}^{hh}$ change sign on another contour that gives the cross-ratio divergence shown in Fig.~\ref{figure_QD_ratio3_3D}.
}
\label{figure_noise_sign}
\end{center} 
\end{figure}

Finally, we propose to further examine the noise ratios in the intermediate regime. Figure~\ref{figure_QD_ratio3_3D} shows these ratios as a function of voltage and temperature gradients without any limitation on their relative values. For this particular QD working, all ratios remain almost insensitive to the temperature gradient while they vary strongly with the voltage. Auto- and cross-ratios are still distinct. Remarkably, $\rLR$ exhibits a divergence at a voltage value for which $\ShhLR$ cancels as shown Fig.~\ref{figure_noise_sign}. 
The sign of the auto-correlators in the right reservoir stays positive whatever the voltage and temperature values, whereas the mixed auto-correlators in the left reservoir show a sign inversion which does not affect the product $\mathcal{S}_{LL}^{eh}\mathcal{S}_{LL}^{he}$. The cross-ratio $\rLR$ changes sign twice: once with $\mathcal{S}_{LR}^{he}$ (see Fig.~\ref{figure_noise_sign}), and the other at a larger voltage due to the change of sign of $\mathcal{S}_{LR}^{hh}$ giving the divergence of $\rLR$. Indeed,  the heat cross-correlator, which is negative at low voltage, becomes positive at high voltage due to the contribution of the term $V^2\mathcal{S}_{LL}^{ee}$ in its expression (see Eq.~(\ref{Shhtot})). 
For a QPC working in the same conditions (not shown here), the charge and heat noises in the same reservoirs, $\SeeLL$ and $\ShhLL$, stay positive whereas the mixed noises, $\SehLL$ and $\SheLL$, can take negative values as for the QD. The charge noise between distinct reservoirs, $\SeeLR$ is negative while its heat counterpart $\ShhLR$ is positive in this regime.
Thus, for the two nanosystems considered here, the results are in agreement with Ref.~\onlinecite{moskalets14} where it has been shown that the heat cross-correlator $\mathcal{S}_{LR}^{hh}$ is not necessary negative, contrary to the charge cross-correlator $\mathcal{S}_{LR}^{ee}$.

%%%%%%%%%%%%%%%%%%%%%%%%%%%%%%%%%%%%%%%%%%%%%%%%%%%%%%%%%%%%%%%%%%
%																 %
%																 %
%		CONCLUSION											 %
%																 %
%																 %
%%%%%%%%%%%%%%%%%%%%%%%%%%%%%%%%%%%%%%%%%%%%%%%%%%%%%%%%%%%%%%%%%%

\section{Conclusion} 

We investigated mixed, charge and heat zero-frequency noises in thermoelectric nanosystems connected to reservoirs using the Landauer-B\"uttiker formalism. 
In the future perspective of studying thermoelectric conversion, we explored two routes. 
On the one hand, we developed relations between the noises and thermoelectric differential conductances which are the adequate quantities to consider in the non-linear regime. 
On the other hand, we interconnected the different noises via ratios of the product of mixed noises divided by the product of charge and heat noises, calculated inside the same reservoir ($\rLL$ and $r_{RR}$) or in between two ($\rLR=r_{RL}$). 
From general derivations, we are able to obtain analytical expressions for differential conductances and noises in various limits. 
The strategy was thus to exploit them in the linear regime of high temperature, and in the non-linear regime of high voltage in two related nanosystems: a quantum point contact and a quantum dot.
Our main conclusions follow.

The mixed conductances $X$ and $Y$ are related to the Seebeck and Peltier coefficients in the linear response regime.
Applying our results to a QPC and a QD, we find that the differential conductances $X$ and $Y$ cancel for systems with electron-hole symmetry. The same applies for the mixed noises $\Sehpq$ and the ratios of noises $\rpq$ in the high temperature regime. 
Inversely, in the non-linear high voltage regime, $X$ and $Y$ still cancel for $\epsilon_0=0$, but neither $\Sehpq$, nor $\rpq$, thus the ratio of noises is no longer related to the ratio of differential conductances in this regime.

The correlations between heat and charge currents provide an indication of the efficiency of thermoelectric conversion in the linear response regime.
Indeed, we have shown that the figure of merit $ZT_0$ is given by the ratio of noises: $ZT_0=S^{eh}_{pp}S^{he}_{pp}/(S^{ee}_{pp}S^{hh}_{pp}-S^{eh}_{pp}S^{he}_{pp})=r_{pp}/(1-r_{pp})$.
We thus have proved from noises calculations that $ZT_0$ is not bounded in that regime since $|r_{pp}|$ can only take value between zero and one.
Moreover, choosing auto-correlations (in the same reservoir), or cross-correlations (between distinct reservoirs), we get a unique expression for the ratios of noises.
This is no longer the case in the high voltage regime where $\rLL$,  $\rRR$ and $\rLR$ take different values: because of the Cauchy-Schwarz inequality, $\rLL$ and $\rRR$ stay smaller than one whereas there is no limitation for $\rLR$. 
The situation is more complex in intermediate regime, where the two auto-ratios $\rLL$ and $\rRR$ are different and show an asymmetry arising from different heat noises in the two reservoirs. 
Moreover, the cross-ratio exhibits a divergence in the QD that occurs when the heat cross-correlation changes sign varying temperature and voltage gradients.

The cross-ratio $\rLR$, introduced for the first time in this paper, deserves to be studied on an equal footing than $\rLL$ and $\rRR$ since it measures how the heat current in one reservoir and the charge current in the other  are related to each other.
  In the case of the QD, we found that the efficiency can be fully  expressed in terms of cross-noises  in the non-linear Schottky regime: $\eta=(\mathcal{S}^{eh}_{LR})^2/|\mathcal{S}^{ee}_{LR}\mathcal{S}^{hh}_{LR}-(\mathcal{S}^{eh}_{LR})^2|$. This result clearly shows that the mixed noise evidences the thermoelectric conversion both in the linear and non-linear regimes. Knowing that the figure of merit is no longer connected to the thermoelectric efficiency in the non-linear regime, there is a need to find a new parameter which informs about the efficiency. These ratios of noises are a possible avenue of research.

\acknowledgments
We thank P.~Eym\'eoud, M.~Guigou and R.~Whitney for their interest in this work and for 
valuable discussions.

%%%%%%%%%%%%%%%%%%%%%%%%%%%%%%%%%%%%%%%%%%%%%%%%%%%%%%%%%%%%%%%%%%
%																 %
%																 %
%		APPENDIX											 %
%																 %
%																 %
%%%%%%%%%%%%%%%%%%%%%%%%%%%%%%%%%%%%%%%%%%%%%%%%%%%%%%%%%%%%%%%%%%

\appendix 

\section{QPC currents and noises in the high voltage regime}

Taking $T_L=T_R=0$, the integrals in the expressions of the currents given by Eqs.~(\ref{def_curr_elec}) and (\ref{def_curr_ther}) can be performed analytically:
\begin{eqnarray}
I^{e}_{L,R}&=&\pm\frac{e}{h}\left[eV-2\Gamma\arctan\left(\frac{eV}{2\Gamma}\right)\right]~,\\
I^{h}_{L,R}&=&\mp\frac{V}{2}I^{e}_{L,R}~.
\end{eqnarray}

The same applies for the expressions of the noises of Eqs.~(\ref{See_zeroT}) to (\ref{Shhcross_zeroT}). We obtain for the auto-correlators:
\begin{eqnarray}
\mathcal{S}^{ee}_{LL}&=&\frac{e^2}{h}\mathrm{sign}(V)\Bigg[-\frac{eV\Gamma^2}{2[\Gamma^2+(eV/2)^2]}\nonumber\\
&&+\Gamma\arctan\left(\frac{eV}{2\Gamma}\right)\Bigg]~,
\end{eqnarray}
\begin{eqnarray}
\mathcal{S}^{eh}_{LL}&=&\frac{e}{h}\mathrm{sign}(V)\Bigg[\frac{(eV\Gamma)^2}{4[\Gamma^2+(eV/2)^2]}\nonumber\\
&&-\frac{eV\Gamma}{2}\arctan\left(\frac{eV}{2\Gamma}\right)\Bigg]~,
\end{eqnarray}
\begin{eqnarray}
\mathcal{S}^{eh}_{RR}&=&\frac{e}{h}\mathrm{sign}(V)\Bigg[-\frac{(eV\Gamma)^2}{4[\Gamma^2+(eV/2)^2]}\nonumber\\
&&+\frac{eV\Gamma}{2}\arctan\left(\frac{eV}{2\Gamma}\right)\Bigg]~,
\end{eqnarray}
\begin{eqnarray}
\mathcal{S}^{hh}_{LL}&=&\frac{1}{h}\mathrm{sign}(V)\Bigg[\frac{eV\Gamma^2}{8}\frac{(eV)^2+12\Gamma^2}{\Gamma^2+(eV/2)^2}\nonumber\\
&&+\frac{\Gamma}{4}[(eV)^2-12\Gamma^2]\arctan\left(\frac{eV}{2\Gamma}\right)\Bigg]~,
\end{eqnarray}
\begin{eqnarray}
\mathcal{S}^{hh}_{RR}&=&\mathcal{S}^{hh}_{LL}~,
\end{eqnarray}

and for the cross-correlators:
\begin{eqnarray}
&&\mathcal{S}^{ee}_{LR}=-\mathcal{S}^{ee}_{LL}~,\\
&&\mathcal{S}^{he}_{LR}=-\mathcal{S}^{eh}_{LL}~,\\
&&\mathcal{S}^{eh}_{LR}=-\mathcal{S}^{eh}_{RR}~,\\
&&\mathcal{S}^{hh}_{LR}=-\frac{1}{h}\mathrm{sign}(V)\Bigg[\frac{3\Gamma^2eV}{2}\nonumber\\
&&-\frac{\Gamma}{4}[(eV)^2+12\Gamma^2]\arctan\left(\frac{eV}{2\Gamma}\right)\Bigg]~.
\end{eqnarray}

\section{QD currents and noises in the high voltage regime}

Performing the integration of Eqs.~(\ref{def_curr_elec}) and (\ref{def_curr_ther}) at zero temperature for the QD, the currents read as:
\begin{eqnarray}
I^{e}_{L,R}&=&\pm\frac{e\Gamma}{h}\sum_\pm\left[\pm\arctan\left(\frac{\epsilon_0\pm eV/2}{\Gamma}\right)\right]~,\\
I^{h}_{L,R}&=&\pm\frac{\Gamma(\epsilon_0\mp eV/2)}{h}\sum_\pm\left[\pm\arctan\left(\frac{\epsilon_0\pm eV/2}{\Gamma}\right)\right]\nonumber\\
&&\pm\frac{\Gamma^2}{2h}\ln\left[\frac{\Gamma^2+(\epsilon_0- eV/2)^2}{\Gamma^2+(\epsilon_0+ eV/2)^2}\right]~.
\end{eqnarray}

In addition, Eqs.~(\ref{See_zeroT}) to (\ref{Shhcross_zeroT}) give at zero temperature for the auto-correlators:
\begin{eqnarray}
\mathcal{S}^{ee}_{LL}&=&\frac{e^2}{h}\mathrm{sign}(V)\sum_\pm \Bigg[\pm\frac{\Gamma^2}{2}\frac{\epsilon_0\mp eV/2}{(\epsilon_0\mp eV/2)^2+\Gamma^2}\nonumber\\
&&\mp\frac{\Gamma}{2}\mathrm{arctan}\left(\frac{\epsilon_0\mp eV/2}{\Gamma}\right) \Bigg]~,
\end{eqnarray}
\begin{eqnarray}
\mathcal{S}^{eh}_{LL}&=&\frac{e}{h}\mathrm{sign}(V)\Bigg(\sum_\pm \Bigg[\frac{\Gamma^2eV}{4}\frac{\epsilon_0\mp eV/2}{(\epsilon_0\mp eV/2)^2+\Gamma^2}\nonumber\\
&&\mp\frac{\Gamma\epsilon_0}{2}\mathrm{arctan}\left(\frac{\epsilon_0\mp eV/2}{\Gamma}\right)\Bigg]\nonumber\\
&&+\frac{\Gamma^2}{2}\ln\left[\frac{\Gamma^2+(\epsilon_0- eV/2)^2}{\Gamma^2+(\epsilon_0+ eV/2)^2}\right]\Bigg)-\frac{V}{2}\mathcal{S}^{ee}_{LL}~,\nonumber\\
\end{eqnarray}
\begin{eqnarray}
\mathcal{S}^{eh}_{RR}&=&\frac{e}{h}\mathrm{sign}(V)\Bigg(\sum_\pm \Bigg[\frac{\Gamma^2eV}{4}\frac{\epsilon_0\mp eV/2}{(\epsilon_0\mp eV/2)^2+\Gamma^2}\nonumber\\
&&\mp\frac{\Gamma\epsilon_0}{2}\mathrm{arctan}\left(\frac{\epsilon_0\mp eV/2}{\Gamma}\right)\Bigg]\nonumber\\
&&+\frac{\Gamma^2}{2}\ln\left[\frac{\Gamma^2+(\epsilon_0- eV/2)^2}{\Gamma^2+(\epsilon_0+ eV/2)^2}\right]\Bigg)+\frac{V}{2}\mathcal{S}^{ee}_{RR}\nonumber\\
&=&\mathcal{S}^{eh}_{LL}+V\mathcal{S}^{ee}_{LL}~,
\end{eqnarray}
\begin{eqnarray}
\mathcal{S}^{hh}_{LL}&=&\frac{\mathrm{sign}(V)}{h}\Bigg(\sum_\pm\Bigg[\frac{\Gamma^2eV}{2}\nonumber\\
&&\pm\frac{\Gamma^2}{2}\frac{\epsilon_0^2(\epsilon_0\mp eV/2)+\Gamma^2(\epsilon_0\pm eV/2)}{(\epsilon_0\mp eV/2)^2+\Gamma^2}\nonumber\\
&&\mp \frac{\Gamma(\epsilon_0^2-3\Gamma^2)}{2}\mathrm{arctan}\left(\frac{\epsilon_0\mp eV/2}{\Gamma}\right)\Bigg]\nonumber\\
&&+\Gamma^2\epsilon_0\ln\left[\frac{\Gamma^2+(\epsilon_0- eV/2)^2}{\Gamma^2+(\epsilon_0+ eV/2)^2}\right]\Bigg)\nonumber\\
&&-V\mathcal{S}^{eh}_{LL}-\frac{V^2}{4}\mathcal{S}^{ee}_{LL}~,
\end{eqnarray}

\begin{eqnarray}
\mathcal{S}^{hh}_{RR}&=&\frac{\mathrm{sign}(V)}{h}\Bigg(\sum_\pm\Bigg[\frac{\Gamma^2eV}{2}\nonumber\\
&&\pm\frac{\Gamma^2}{2}\frac{\epsilon_0^2(\epsilon_0\mp eV/2)+\Gamma^2(\epsilon_0\pm eV/2)}{(\epsilon_0\mp eV/2)^2+\Gamma^2}\nonumber\\
&&\mp \frac{\Gamma(\epsilon_0^2-3\Gamma^2)}{2}\mathrm{arctan}\left(\frac{\epsilon_0\mp eV/2}{\Gamma}\right)\Bigg]\nonumber\\
&&+\Gamma^2\epsilon_0\ln\left[\frac{\Gamma^2+(\epsilon_0- eV/2)^2}{\Gamma^2+(\epsilon_0+ eV/2)^2}\right]\Bigg)\nonumber\\
&&+V\mathcal{S}^{eh}_{RR}-\frac{V^2}{4}\mathcal{S}^{ee}_{RR}\nonumber\\
 &=&\mathcal{S}^{hh}_{LL}+2V\mathcal{S}^{eh}_{LL}+V^2\mathcal{S}^{ee}_{LL}~,
\end{eqnarray}

and for the cross-correlators:
\begin{eqnarray}
&&\mathcal{S}^{ee}_{LR}=-\mathcal{S}^{ee}_{LL}~,
\end{eqnarray}
\begin{eqnarray}
&&\mathcal{S}^{he}_{LR}=-\mathcal{S}^{eh}_{LL}~,
\end{eqnarray}
\begin{eqnarray}
&&\mathcal{S}^{eh}_{LR}=-\mathcal{S}^{eh}_{RR}~,
\end{eqnarray}
\begin{eqnarray}
&&\mathcal{S}^{hh}_{LR}=-\frac{\mathrm{sign}(V)}{h}\Bigg(\sum_\pm\Bigg[\frac{\Gamma^2eV}{2}\nonumber\\
&&\pm\frac{\Gamma^2}{2}\frac{\epsilon_0^2(\epsilon_0\mp eV/2)+\Gamma^2(\epsilon_0\pm eV/2)}{(\epsilon_0\mp eV/2)^2+\Gamma^2}\nonumber\\
&&\mp \frac{\Gamma(\epsilon_0^2-3\Gamma^2)}{2}\mathrm{arctan}\left(\frac{\epsilon_0\mp eV/2}{\Gamma}\right)\Bigg]\nonumber\\
&&+\Gamma^2\epsilon_0\ln\left[\frac{\Gamma^2+(\epsilon_0- eV/2)^2}{\Gamma^2+(\epsilon_0+ eV/2)^2}\right]\Bigg)-\frac{V^2}{4}\mathcal{S}^{ee}_{LR}~.\nonumber\\
\end{eqnarray}

%%%%%%%%%%%%%%%%%%%%%%%%%%%%%%%%%%%%%%%%%%%%%%%%%%%%%%%%%%%%%%%%%%
%																 %
%																 %
%		REFERENCES											 %
%																 %
%																 %
%%%%%%%%%%%%%%%%%%%%%%%%%%%%%%%%%%%%%%%%%%%%%%%%%%%%%%%%%%%%%%%%%%


\begin{thebibliography}{99}
\bibitem{note1}In reality, the transmission coefficient depends on the voltage through the density of states of the reservoirs. However, for voltage smaller than the band width, it is justified to use a voltage independent transmission (wide-band limit). At voltage higher compared to the band width, this assumption is no longer justified and additional non-linear contributions to the currents and noises should appear.
\bibitem{note2}It is important to point out that at voltage higher compared to the band width, the transmission coefficient acquires a voltage dependence which breaks the electron-hole symmetry and gives non-zero $X$ and $Y$ differential conductances. We thus assume here that the voltage stays smaller than the band width.
\end{thebibliography}
\end{document}